\documentclass[a4paper,11pt]{article}

\usepackage{jheppub}
\usepackage{amsmath,amssymb,amscd}
\usepackage{bbm}
\usepackage{graphicx}
\usepackage{url}

\graphicspath{ {fig/} }

\makeatletter
\def\hlinewd#1{%
  \noalign{\ifnum0=`}\fi\hrule \@height #1 \futurelet
   \reserved@a\@xhline}
\makeatother

\makeatletter
\renewcommand\@fpheader{}
\renewcommand\@journal{}
\makeatother

\newcommand{\Graph}[2][0.3]{\vcenter{\hbox{\includegraphics[scale=#1]{#2}}}}
\newcommand{\dimgraph}[2]{\overset{(#2-2\epsilon)}{\Graph[0.2]{#1}}}
\newcommand{\figgraph}[3]{\overset{(#3-2\epsilon)}{\Graph[#1]{#2}}}
\definecolor{darkgreen}{rgb}{0.,.3,0}
\definecolor{darkblue}{rgb}{0.0,0.0,0.5}
\newcommand{\pole}[1]{\textcolor{darkblue}{#1}}
\newcommand{\casimir}[1]{\textcolor{darkgreen}{#1}}
\newcommand{\HyperInt}{\texttt{HyperInt}}
\newcommand{\file}[1]{\texttt{#1}}
\newcommand{\Maple}{\texttt{Maple}}

\title{
On the Computation of Form Factors in \\Massless QCD with Finite Master Integrals
}

\preprint{MITP/15-072, TCDMATH-15-10}

\author[a]{Andreas von Manteuffel,}
\author[b]{Erik Panzer,}
\author[\,c]{and Robert M.~Schabinger}

\affiliation[a]{PRISMA Cluster of Excellence,
Institut f\"ur Physik,
Johannes-Gutenberg Universit\"{a}t,\\
55099 Mainz, Deutschland}
\affiliation[b]{All Souls College, Oxford, OX1 4AL, United Kingdom}
\affiliation[c]{Hamilton Mathematical Institute, Trinity College,
Dublin 2, Ireland}

\emailAdd{manteuffel@uni-mainz.de}
\emailAdd{erik.panzer@all-souls.ox.ac.uk}
\emailAdd{schabr@maths.tcd.ie}

\abstract{%
We present the bare one-, two-, and three-loop form factors in massless
Quantum Chromodynamics as linear combinations of finite master integrals.
Using symbolic integration, we compute their $\epsilon$ expansions and
thereby reproduce all known results with an independent method.
Remarkably, in our finite basis, only integrals with a less-than-maximal
number of propagators contribute to the cusp anomalous dimensions.
We report on indications of this phenomenon at four loops, including the
result for a finite, irreducible, twelve-propagator form factor integral.
Together with this article, we provide our automated software setup for
the computation of finite master integrals.%
}

\begin{document}
\unitlength1cm
\maketitle
\allowdisplaybreaks

\section{Introduction} 
\label{sec:intro}
In recent years, higher-order perturbative calculations of the quark and gluon form factors in massless Quantum Chromodynamics (QCD) have generated a great deal of interest. 
This is primarily due to their relevance to two of the most important Large Hadron Collider processes, the production of a Drell-Yan lepton pair \cite{Drell:1970wh}
and the production of a Higgs boson via gluon fusion \cite{Georgi:1977gs}. 
At the most basic level, a calculation of the quark and gluon form factors at some fixed order provides, respectively, the purely virtual QCD corrections to Drell-Yan lepton production 
and gluon-fusion Higgs boson production in the infinite top-quark-mass limit~\cite{Wilczek:1977zn,Shifman:1978zn,Ellis:1979jy,Inami:1982xt}.
Furthermore, the cusp and collinear anomalous dimensions can be extracted from $\epsilon^{-2}$ and $\epsilon^{-1}$ poles of the bare form factors.
They determine the structure of the quark and gluon jet functions, objects which play an important role in the theory of the infrared divergences of multi-leg, massless scattering amplitudes
and in the theory of soft-gluon resummation (see {\it e.g.} reference \cite{Becher:2014oda} for a recent review).

Even if all real radiation is very soft relative to the scale of the hard scattering, the purely virtual corrections to a process are only one part of the story.
Using the eikonal approximation to treat all real radiation allows one to give an approximate cross section for a process at higher orders in QCD perturbation theory.
In reference \cite{Harlander:2002wh}, this so-called soft-virtual approximation was shown
to capture a sizable fraction of the total gluon-fusion Higgs boson production cross section at next-to-next-to-leading order.
Although three-loop calculations of both form factors through the finite terms \cite{Baikov:2009bg,Lee:2010cga,Gehrmann:2010ue} were key ingredients for recent
third-order, soft-virtual calculations of the cross sections for Drell-Yan lepton production and gluon-fusion Higgs boson production, much more work is required to obtain the final results. 
As usual, it is also necessary to consider the radiation of additional partons and then combine all relevant real radiative
corrections with the purely virtual contributions. Only then are infrared-finite results obtained which can be convoluted with the appropriate parton distribution functions to produce meaningful numbers.
In a series of papers \cite{Anastasiou:2013srw,Li:2013lsa,Duhr:2013msa,Anastasiou:2014vaa,Li:2014bfa,Ahmed:2014cla,Li:2014afw}, results were obtained which,
in principle, could have determined both the Drell-Yan lepton production cross section and the gluon-fusion Higgs boson production cross section to levels of precision sufficient for the purposes of LHC physics.
All of this work culminated in a new milestone, namely two independent calculations of the parton-level cross sections for both processes in the soft-virtual
next-to-next-to-next-to-leading order (N$^3$LO) approximation \cite{Anastasiou:2014vaa,Ahmed:2014cla,Li:2014afw}. 

A systematic inclusion of power-corrections in the soft-virtual Higgs-production
analysis~\cite{Anastasiou:2014lda} shows significant contributions beyond the
leading term. While the small numerical impact at high expansion order suggests that this approximate N$^3$LO prediction is sufficiently precise for the purposes of Large Hadron Collider phenomenology,
it is still interesting to carry out the exact N$^3$LO calculation to put the analysis on more rigorous grounds.
It has long been known that the accuracy of predictions for the above-mentioned production processes can be improved significantly if appropriate towers of logarithms 
coming from soft radiation near the production threshold are resummed~\cite{Kramer:1996iq}.
To improve upon the current state-of-the-art, one could carry out a soft parton resummation of the next-to-next-to-next-to-leading Sudakov logarithms (N$^3$LL order) 
and then match this onto the exact fixed order N$^3$LO prediction once the result becomes available.
At this stage, it would also be of interest 
to include a resummation of additional terms in the virtual matrix elements which appear due to the fact that $s$-channel processes have a time-like momentum transfer \cite{Ahrens:2008qu,Ahrens:2008nc}.

In fact, some partial progress towards this more ambitious goal has already been realized.
Given the impressive number of recently-obtained 
results \cite{Hoschele:2012xc,Anastasiou:2013mca,Kilgore:2013gba,Hoschele:2014qsa,Duhr:2014nda,Dulat:2014mda,Anastasiou:2015ema,Anastasiou:2015yha,Bonvini:2014joa,Catani:2014uta,deFlorian:2014vta,Anzai:2015wma}, 
it seems clear that the exact parton-level N$^3$LO results lie within reach and that,
furthermore, matching them with N$^3$LL Sudakov resummations will not be an issue once all of the required ingredients
become available. Actually, almost all of the quantities required for a complete N$^3$LL resummation have already been available in the literature for quite some time, both for Drell-Yan and for Higgs. However, in order to properly carry out a 
N$^3$LL resummation, one needs the four-loop cusp anomalous dimensions. Although the three-loop cusp anomalous dimensions were calculated long ago \cite{Moch:2004pa},
surprisingly little has been reported at one order higher; several interesting techniques have been 
developed \cite{Panzer:2013cha,Panzer:2014gra,Panzer:2014caa,Panzer:2015ida,vonManteuffel:2014qoa,vonManteuffel:2014ixa,Ablinger:2012ph,Ablinger:2015tua,Ruijl:2014hha,Lee:2012te,Henn:2013nsa}
and a number of relevant master integrals (mainly propagators\footnote{One four-loop three-point integral with off-shell external momenta was presented in \cite[example~3.6]{Panzer:2014gra}.}) computed, 
see \cite{Baikov:2010hf,Smirnov:2010hd} and \cite{Lee:2011jf,Lee:2011jt}\footnote{Recently, it was reported that these computations can now be reproduced using the public software package
{\tt SummerTime} \cite{Lee:2015eva}.}, but it is arguably the case that more progress has been made on the analogous problem
in maximally supersymmetric gauge theory \cite{Boels:2012ew,Boels:2015yna}.

Besides the phenomenological motivation given above, a calculation of the four-loop cusp anomalous dimensions also has the potential to answer a long-standing question about the infrared structure of massless QCD. 
Through three-loop order, it has been observed that the gluon cusp anomalous dimension can be derived from the quark cusp anomalous dimension by simply rescaling it. 
The calculation of both the quark and the gluon cusp anomalous dimensions at four loops is therefore essential to see if this pattern continues to hold. 
These four-loop calculations will provide the first non-trivial test of the putative Casimir scaling property of QCD:
the proposal that, to all loop orders in massless QCD, the gluon cusp anomalous dimension is the same as the quark cusp anomalous dimension up to an overall factor of $C_A/C_F$, where $C_A = N_c$ and $C_F = (N_c^2-1)/(2 N_c)$
are respectively the quadratic Casimir invariants of the adjoint and the fundamental representations of the QCD gauge group.\footnote{We assume a $SU(N_c)$ gauge group throughout this article.} 
A breakdown of Casimir scaling at four loop order would have profound implications. For example, it would mean that the four-loop infrared divergences in massless gauge theory scattering amplitudes at subleading color
cannot straightforwardly be described in an abstract way which treats quarks and gluons on an equal footing \cite{Becher:2009qa,Gardi:2009qi}.

In this paper we demonstrate the advantages of our approach \cite{vonManteuffel:2014qoa} in a complete rederivation of the well-studied one-, two-, and three-loop form factors in perturbative QCD. 
We found it particularly elegant to use a suitable basis of finite master integrals and made three main observations:
\begin{itemize}
	\item
The $\epsilon$ pole structure of the unexpanded form factors becomes absolutely explicit and has the striking feature that the most complicated integrals do not contribute to the cusp anomalous dimensions.

	\item
Finite integrals can be computed exactly and automatically in terms of multiple polylogarithms. In fact, our work constitutes the first complete analytical check of the weight eight, 
$\mathcal{O}\left(\epsilon^2\right)$ three-loop results published in \cite{Gehrmann:2010tu}.

	\item
On the numerical side, we obtained more than an order of magnitude improvement in both run time and precision for a fixed number of Monte Carlo integrand evaluations by employing a basis of finite integrals.
Remarkably, we have since observed that, quite generally, one can substantially increase the reach and reliability of publicly available sector decomposition programs \cite{Binoth:2000ps,Bogner:2007cr,Smirnov:2013eza,Borowka:2015mxa} 
if one first rotates to a basis of finite integrals. We will explore this in detail in a separate publication.
\end{itemize}
Towards the end of this work, we take a first look at the computation of the four-loop cusp anomalous dimensions and explain how our method will allow for a significant reduction of the problem. 
We illustrate our point at four loops by computing a non-trivial integral in an irreducible top-level sector. 
Due to the fact that the leading term in its $\epsilon$ expansion consists solely of transcendental numbers of weight seven, it is not expected to contribute to the cusp anomalous dimensions.

This article is organized as follows. In Section~\ref{sec:notation}, we define our $L$-loop bare form factors in massless QCD, introduce notation that we use,
and state some useful facts about the structure of the form factors at one, two, and three loops. 
In Section~\ref{sec:method}, we describe our method of computation, focusing on its non-standard features.
In Sections \ref{sec:1Lffs}, \ref{sec:2Lffs}, and \ref{sec:3Lffs}, we present our results for the one-, two- and three-loop bare form factors written as linear combinations of finite master integrals. 
When written in this way, the results have the remarkable feature that many of the most complicated master integrals do not contribute to the $\epsilon^{-2}$ pole terms. 
We explore this further in Section~\ref{sec:cusps} and give an example of this phenomenon at the four-loop level,
based on the exact computation of a four-loop form factor with {\HyperInt}, a package for the evaluation of convergent Feynman integrals \cite{Panzer:2014caa}.
In Appendix~\ref{sec:finite}, we extend theoretical arguments given in reference \cite{vonManteuffel:2014qoa}, showing that,
for scattering and decay processes which admit a Euclidean region respecting the kinematical constraints, one can always find a basis of finite loop integrals.

As a crucial supplement, ancillary files are available on arXiv.org which contain the quark and gluon form factors in terms of finite master integrals at one, two, and three loops,
as well as all of our finite master integrals and final results $\epsilon$-expanded to weight eight.
We also provide our highly-automated and parallelized setup for the computation of the $\epsilon$-expansions using {\HyperInt} as described in Appendix~\ref{sec:hyperint}.
Given sufficient computer resources, all of the Feynman integrals discussed in this paper may be straightforwardly reproduced by following the instructions given in the ancillary files.

\section{Notation and Conventions}
\label{sec:notation}
\begin{figure}
\centering
\includegraphics[scale=0.35]{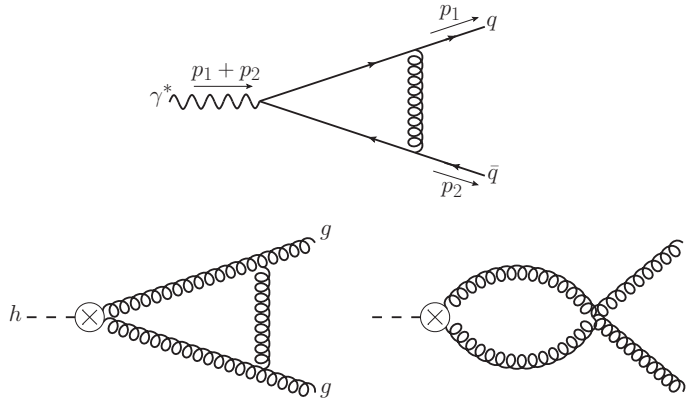}
\caption{The one-loop Feynman diagrams for the quark (upper panel) and gluon (lower panel) form factors in massless QCD. 
At each order in the bare strong coupling constant, the effective coupling of the Higgs boson to gluons can be obtained by matching full QCD with a massive top quark onto an effective field theory in which the massive top quark is integrated out.}
\label{fig:samplediagrams}
\end{figure}

In this section, we give our definitions of the unrenormalized, massless quark and gluon form factors, $\mathcal{F}_{\rm bare}^q\left(\alpha_s^{\rm bare}, (p_1 + p_2)^2, \mu_\epsilon^2, \epsilon\right)$ 
and $\mathcal{F}_{\rm bare}^g\left(\alpha_s^{\rm bare}, (p_1 + p_2)^2, \mu_\epsilon^2, \epsilon\right)$. 
We also establish some notation and state some facts about the general structure of our bare results at one, two, and three loops.
The $L$-loop contribution to the quark form factor is defined by the interference of the tree-level amplitude for $\gamma^*(p_1+p_2) \to q(p_1) \bar{q}(p_2)$ with the $L$-loop corrections to this amplitude in massless QCD, 
normalized to the tree-level amplitude squared. For all interferences, the sum over spin and color degrees of freedom is implied.
Similarly, the $L$-loop contribution to the gluon form factor is defined by the interference of the tree-level diagram for $h(p_1+p_2) \to g(p_1) g(p_2)$ in the infinite top quark mass limit 
with the $L$-loop corrections to this process in massless QCD, normalized to the tree-level amplitude squared (see Figure \ref{fig:samplediagrams} for a visualization at the one-loop level). 

Our bare form factors have a perturbative expansion of the form
\begin{align}
\label{eq:expbareq}
\mathcal{F}_{\rm bare}^q\left(\alpha_s^{\rm bare}, (p_1 + p_2)^2, \mu_\epsilon^2, \epsilon\right) &= 1 + \sum_{L = 1}^\infty \left(\frac{\alpha_s^{\rm bare}}{4\pi}\right)^L 
\left(\frac{4\pi \mu_\epsilon^2}{-(p_1+p_2)^2}\right)^{L \epsilon} \frac{\mathcal{F}_L^q(\epsilon)}{\Gamma^L(1-\epsilon)}
\\
\label{eq:expbareg}
\mathcal{F}_{\rm bare}^g\left(\alpha_s^{\rm bare}, (p_1 + p_2)^2, \mu_\epsilon^2, \epsilon\right) &= 1 + \sum_{L = 1}^\infty \left(\frac{\alpha_s^{\rm bare}}{4\pi}\right)^L 
\left(\frac{4\pi \mu_\epsilon^2}{-(p_1+p_2)^2}\right)^{L \epsilon} \frac{\mathcal{F}_L^g(\epsilon)}{\Gamma^L(1-\epsilon)},
\end{align}
where $\alpha_s^{\rm bare}$ is the bare strong coupling constant, $(p_1 + p_2)^2$ is the momentum transfer squared, $\mu_\epsilon$ is the 't Hooft scale, and
$\epsilon$ is the parameter of dimensional regularization \cite{'tHooft:1972fi}.
The dependence on the scale $(p_1+p_2)^2$ is trivial and may be reconstructed at any time by power counting. We therefore set
\begin{equation}
(p_1+p_2)^2 = -1
\end{equation}
to simplify our notation.

The master integrals for the form factors have traditionally been calculated in $d=4-2\epsilon$ dimensions.
In the present paper, however, we make extensive use of dimensionally-shifted basis integrals.
We employ Minkowskian propagators and choose an absolute normalization of
\begin{equation}
\left(\frac{\Gamma(d/2 - 1)}{i \pi^{d/2}}\right)^{L}
\end{equation}
for our $L$-loop Feynman integrals defined in $d$ dimensions in order to prevent the appearance of spurious constants in our results.
We also consider propagators of higher multiplicity
in our scalar Feynman integrals.
These we visualize by placing dots on the edges of the associated graphical representations,
where a propagator power $\nu + 1$ is represented by an edge with $\nu$ dots on it.
As a concrete example, let us consider the one-loop Feynman integral which we actually use in Section \ref{sec:1Lffs} below. It is the one-loop bubble integral in $d=6 - 2\epsilon$ with both propagators squared:
\begin{align}\label{eq:miff1a3}
	\dimgraph{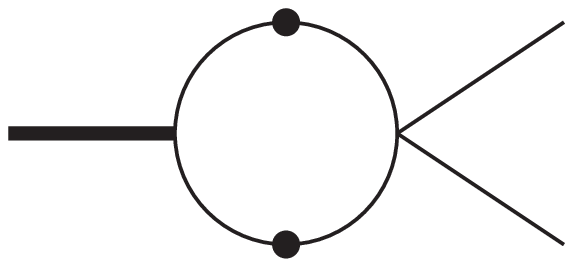}{6} 
	&= 
	\left. \frac{\Gamma(2-\epsilon)}{i \pi^{3-\epsilon}}\int\!\!\frac{\mathrm{d}^{6-2\epsilon}k_1}{((k_1+p_1)^2)^2((k_1-p_2)^2)^2} ~\right|_{(p_1+p_2)^2 = -1}
	\nonumber\\
	&= \frac{\Gamma(2-\epsilon)\Gamma^2(1-\epsilon)\Gamma(1+\epsilon)}{\Gamma(2-2\epsilon)}
	\nonumber\\
	&= 1 + \epsilon + 2 \epsilon^2 + \mathcal{O}\left(\epsilon^3\right).
\end{align}

In Tables \ref{tab:ff1fams}, \ref{tab:ff2fams}, and \ref{tab:ff3fams} below, we present the integral families that we used to uniquely parametrize all of our Feynman integrals. 
In these tables, the loop momenta are denoted by $k_i$, $i=1,2,3$, and each momentum $q$ in the lists represents a scalar propagator $1/q^2$.
At one-loop, the form factors are extremely simple and the three non-zero diagrams of Figure \ref{fig:samplediagrams} can be covered with just one integral family, called $\mathrm{A_1}$ in this work, see Table \ref{tab:ff1fams}.
The color structure at one-loop is also extremely simple; $\mathcal{F}_1^q(\epsilon)$
is directly proportional to $C_F$ and $\mathcal{F}_1^g(\epsilon)$ is directly proportional to $C_A$. 
The two-loop form factors, on the other hand, are more complicated and have an interesting history.
It took three attempts to correctly calculate the two-loop quark form factor through the finite terms \cite{Gonsalves:1983nq,Kramer:1986sg,Matsuura:1987wt}, 
and the analogous calculation for the two-loop gluon form factor was first published more than a decade later in reference \cite{Harlander:2000mg}. 
In reference \cite{Gehrmann:2005pd}, the two-loop form factors were finally computed to all orders in $\epsilon$ in terms of Gamma functions and generalized hypergeometric functions. One must use two integral families
to cover the two-loop Feynman diagrams, chosen as $\mathrm{A_2}$ and $\mathrm{B_2}$ in this paper, see Table \ref{tab:ff2fams}. Both $\mathcal{F}_2^q(\epsilon)$ and $\mathcal{F}_2^g(\epsilon)$ have three color structures,
some of which depend on the number of massless quarks, $N_f$. The color structures $C_F^2$, $C_F C_A$, $C_F N_f$ appear in the expression for $\mathcal{F}_2^q(\epsilon)$
and the color structures $C_A^2$, $C_A N_f$, and $C_F N_f$ appear in the expression for $\mathcal{F}_2^g(\epsilon)$.

Needless to say, the calculation of the three-loop form factors is harder still. To obtain even approximate results at $\mathcal{O}\left(\epsilon^0\right)$ 
took many years of work by a large number of researchers \cite{Moch:2005tm,Gehrmann:2006wg,Heinrich:2007at,Heinrich:2009be,Baikov:2009bg}. Analytical
results for the three-loop form factors through the finite terms were obtained a short time later \cite{Lee:2010cga,Gehrmann:2010ue}. 
The three-loop form factor master integrals were first computed numerically to $\mathcal{O}\left(\epsilon^2\right)$ in reference \cite{Lee:2010ik} using dimensional recurrence relations \cite{Lee:2009dh} but, 
with the help of the celebrated PSLQ algorithm \cite{PSLQ}, the authors were able to recover the analytical solutions.\footnote{At the time, it was conjectured that zeta and multiple zeta values of, at most, 
weight eight would appear in the higher-order results required to expand the three-loop form factors through to $\mathcal{O}\left(\epsilon^2\right)$.} 
The explicit higher-order results for the three-loop masters are, regrettably, scattered over three different articles \cite{Gehrmann:2005pd,Lee:2010ug,Lee:2010ik}. 
Nevertheless, it should be stressed that reference \cite{Lee:2010ik} provided the bulk of the unknown higher-order terms in the $\epsilon$ expansions of the master integrals
required for a calculation of the three-loop form factors up to and including contributions of $\mathcal{O}\left(\epsilon^2\right)$ \cite{Gehrmann:2010tu}. 
Some time later, a subset of the higher-order results for the three-loop masters were confirmed by first solving an auxiliary system of differential equations for analogous integrals with two off-shell legs 
and then performing an asymptotic analysis on the results obtained \cite{Henn:2013nsa}. Three integral families are needed to cover all three-loop Feynman diagrams which contribute. 
The integral families we use coincide with those of reference \cite{Gehrmann:2010ue} and are labeled $\mathrm{A_3}$, $\mathrm{B_3}$, $\mathrm{C_3}$ in Table \ref{tab:ff3fams}. 
The color structures $C_F^3$, $C_F^2 C_A$, $C_F C_A^2$, $C_F^2 N_f$, $C_F C_A N_f$, $C_F N_f^2$, and $(d_{abc}d_{abc}/N_c) N_{q\gamma}$ appear in the expression for $\mathcal{F}_3^q(\epsilon)$ 
and the color structures $C_A^3$, $C_A^2 N_f$, $C_A C_F N_f$, $C_F^2 N_f$, $C_A N_f^2$, and $C_F N_f^2$ appear 
in the expression for $\mathcal{F}_3^g(\epsilon)$. Here, $N_{q\gamma} =  (1/e_q){\sum_{q^\prime} e_{q^\prime}}$ is the charge-weighted sum of the $N_f$ quark flavors normalized to the charge of the primary quark $q$
and $d_{abc}d_{abc} = (N_c^2 - 1)(N_c^2 - 4)/N_c$ for $SU(N_c)$.

\begin{table}
\centering
\begin{tabular}[h!]{l}
Family $\mathrm{A_1}$\\[1mm]
\hlinewd{2pt}
\rule{0pt}{2ex}~~~$k_1+p_1$\\
\rule{0pt}{2ex}~~~$k_1-p_2$\\
\rule{0pt}{2ex}~~~$k_1$
\end{tabular}
\caption{One integral family which covers all one-loop form factor diagrams.}
\label{tab:ff1fams}
\end{table}
\begin{table}
\centering
\begin{tabular}[h!]{ll}
Family $\mathrm{A_2}$\hspace{5mm} & Family $\mathrm{B_2}$\hspace{5mm}\\[1mm]
\hlinewd{2pt}
\rule{0pt}{2ex}~$k_1+p_1$ & $k_1+p_1$\\
\rule{0pt}{2ex}~$k_2+p_1$ & $k_2+p_1$\\
\rule{0pt}{2ex}~$k_1-p_2$ & $k_1-p_2$\\
\rule{0pt}{2ex}~$k_2-p_2$ & $k_1-k_2-p_2$\\
\rule{0pt}{2ex}~$k_1-k_2$ & $k_1-k_2$\\
\rule{0pt}{2ex}~$k_1$     & $k_2$\\
\rule{0pt}{2ex}~$k_2$     & $k_1$
\end{tabular}
\caption{Two integral families which cover all two-loop form factor diagrams.}
\label{tab:ff2fams}
\end{table}
\begin{table}
\centering
\begin{tabular}[h!]{lll}
Family $\mathrm{A_3}$\hspace{5mm} & Family $\mathrm{B_3}$\hspace{5mm} & Family $\mathrm{C_3}$\hspace{5mm}\\[1mm]
\hlinewd{2pt}
\rule{0pt}{2ex}~$k_1$         & $k_1$         & $k_1$\\
\rule{0pt}{2ex}~$k_2$         & $k_2$         & $k_2$\\
\rule{0pt}{2ex}~$k_3$         & $k_3$         & $k_3$\\
\rule{0pt}{2ex}~$k_1-k_2$     & $k_1-k_2$     & $k_1-k_2$\\
\rule{0pt}{2ex}~$k_1-k_3$     & $k_1-k_3$     & $k_1-k_3$\\
\rule{0pt}{2ex}~$k_2-k_3$     & $k_1-k_2-k_3$ & $k_2-k_3$\\
\rule{0pt}{2ex}~$k_1-p_1$     & $k_1-p_1$     & $k_1-k_3-p_2$\\
\rule{0pt}{2ex}~$k_1-p_1-p_2$ & $k_1-p_1-p_2$ & $k_1-p_1-p_2$\\
\rule{0pt}{2ex}~$k_2-p_1$     & $k_2-p_1$     & $k_2-p_1$\\
\rule{0pt}{2ex}~$k_2-p_1-p_2$ & $k_2-p_1-p_2$ & $k_1-k_2-p_2$\\
\rule{0pt}{2ex}~$k_3-p_1$     & $k_3-p_1$     & $k_3-p_1$\\
\rule{0pt}{2ex}~$k_3-p_1-p_2$ & $k_3-p_1-p_2$ & $k_3-p_1-p_2$
\end{tabular}
\caption{Three integral families which cover all three-loop form factor diagrams.}
\label{tab:ff3fams}
\end{table}

\section{Computational Method}
\label{sec:method}
\begin{figure}
\centering
\begin{align*}
&\figgraph{.35}{ff1a_2_3}{6}
\\
&\figgraph{.35}{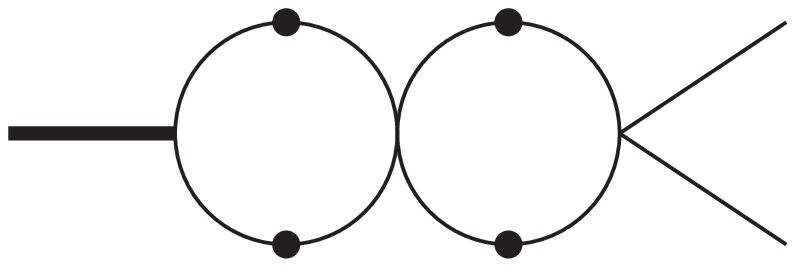}{6} \quad \figgraph{.35}{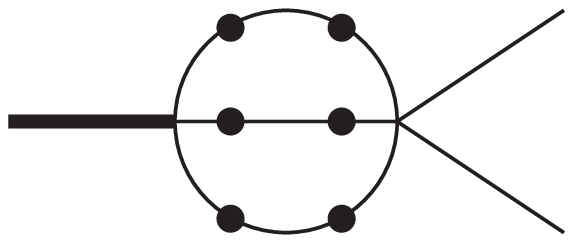}{8} \quad \figgraph{.35}{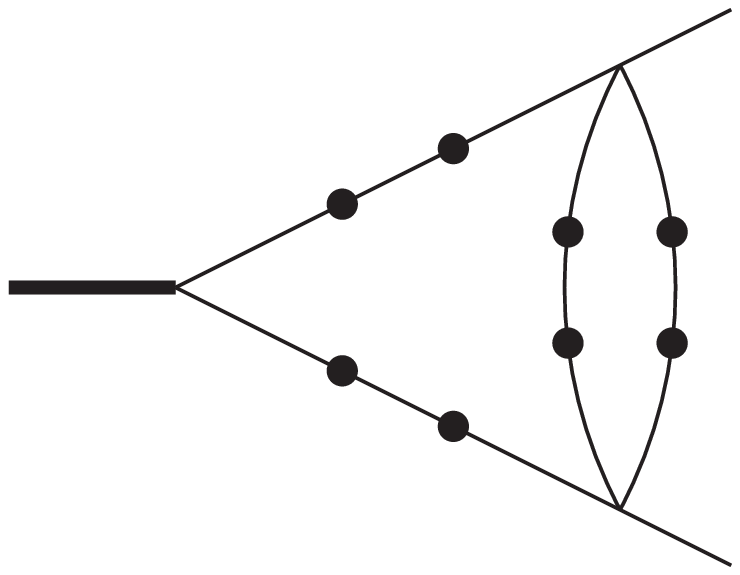}{10} \quad \figgraph{.35}{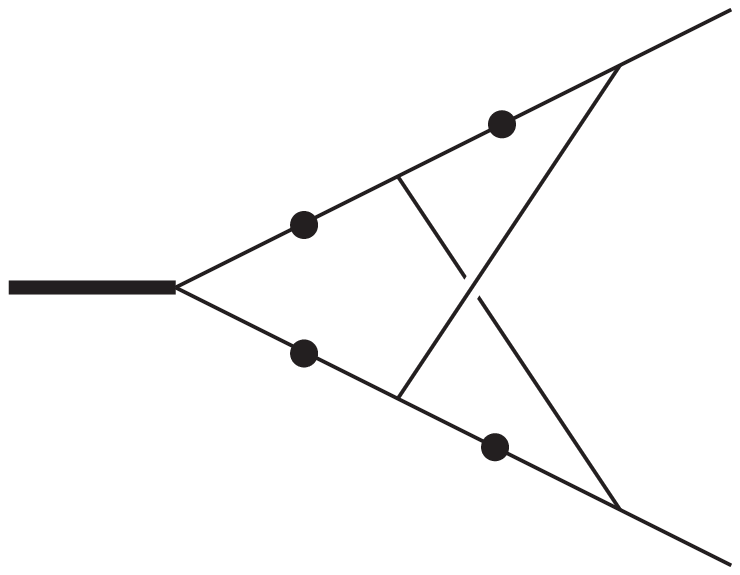}{8}
\\
&\figgraph{.35}{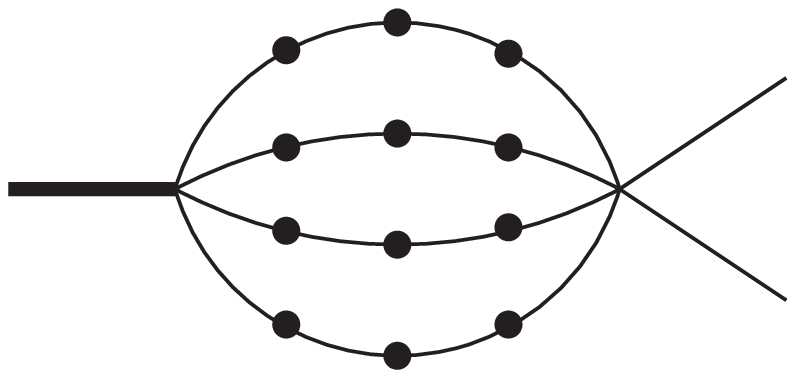}{10} \quad \figgraph{.35}{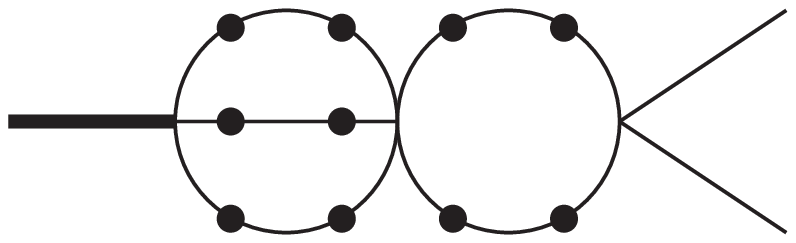}{8} \quad \figgraph{.35}{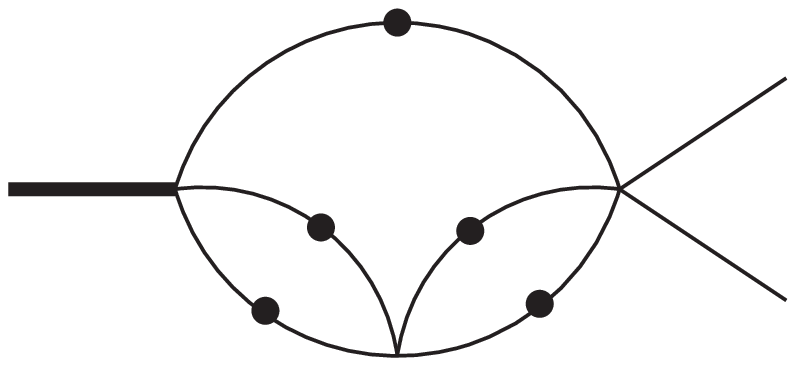}{6} \quad \figgraph{.35}{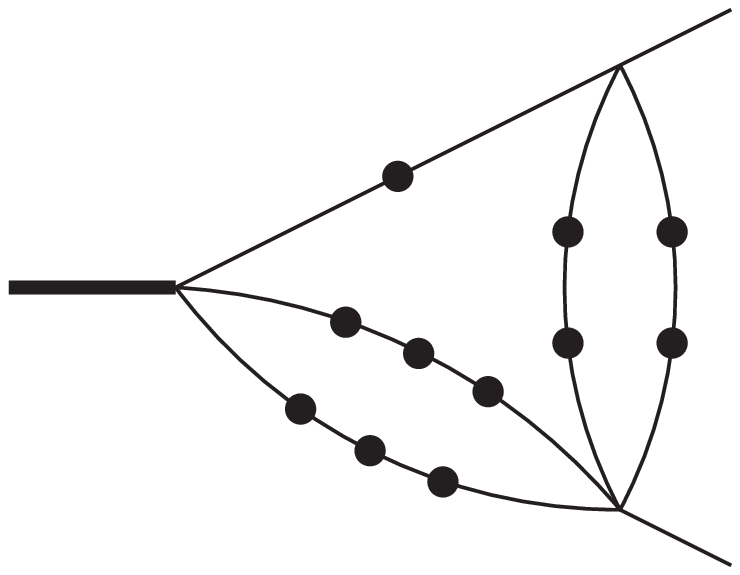}{10}
\\
&\figgraph{.35}{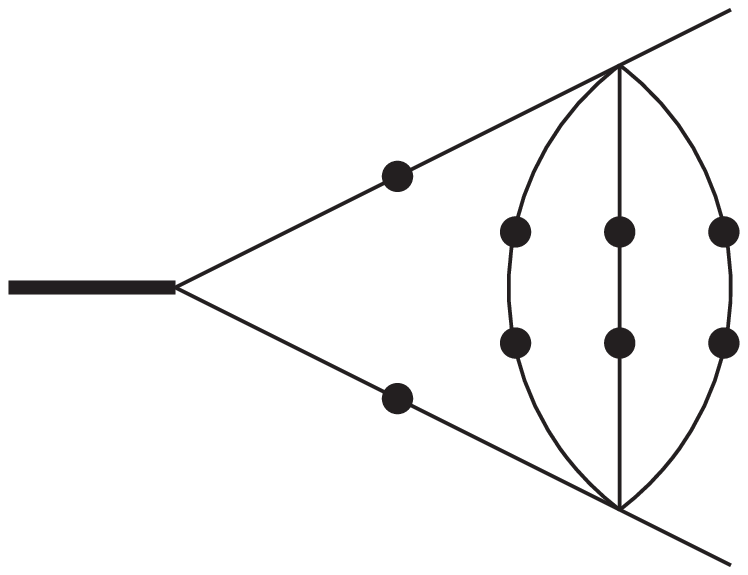}{8} \quad \figgraph{.35}{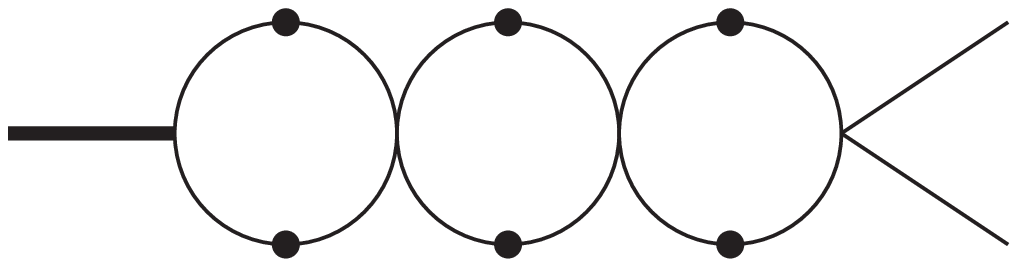}{6} \quad \figgraph{.35}{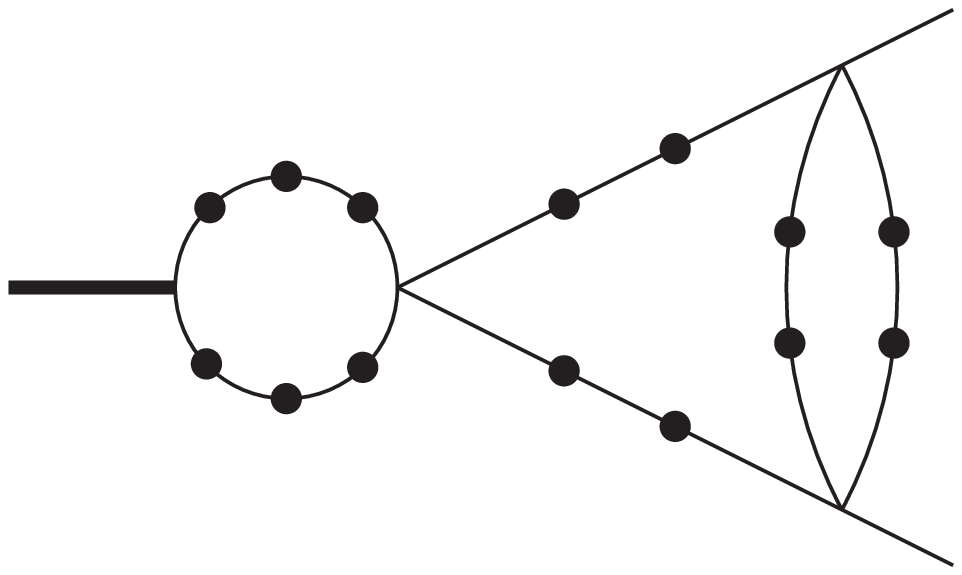}{10} \quad \figgraph{.35}{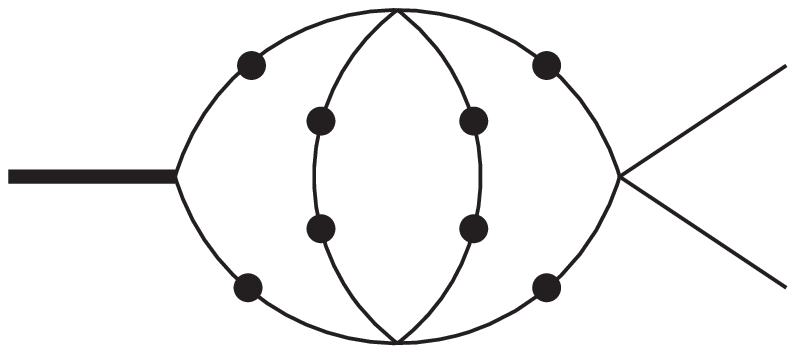}{8}
\\
&\figgraph{.35}{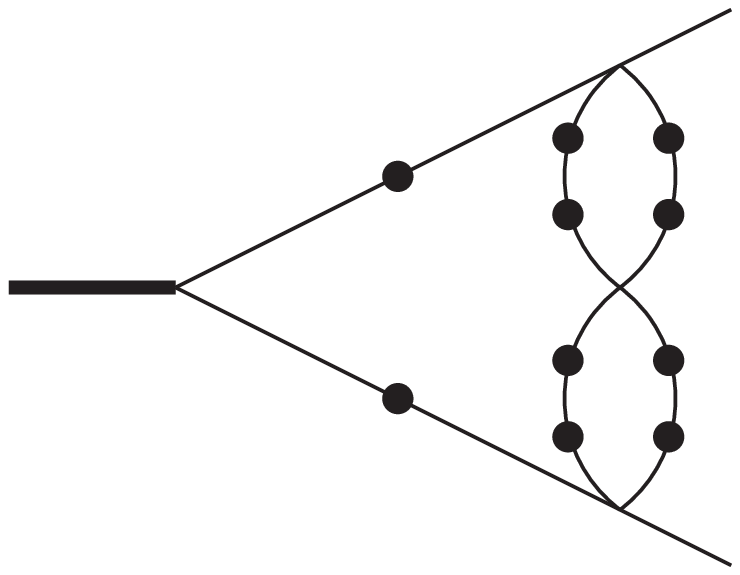}{10} \quad \figgraph{.35}{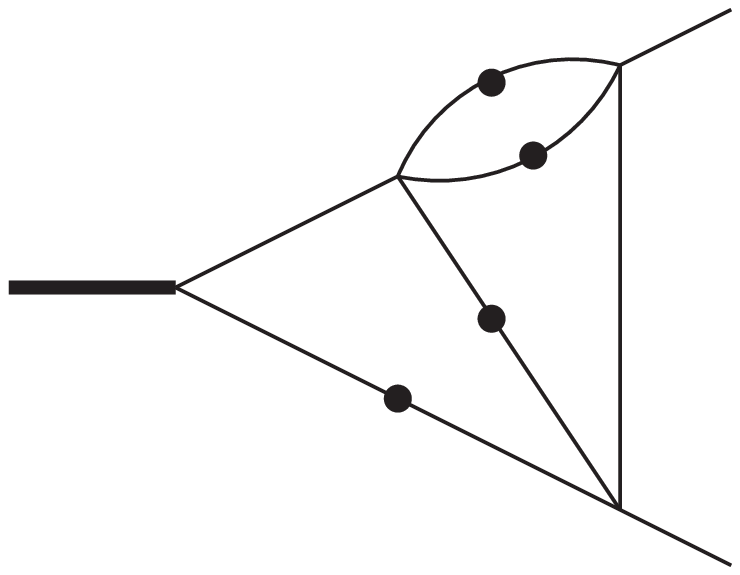}{6} \quad \figgraph{.35}{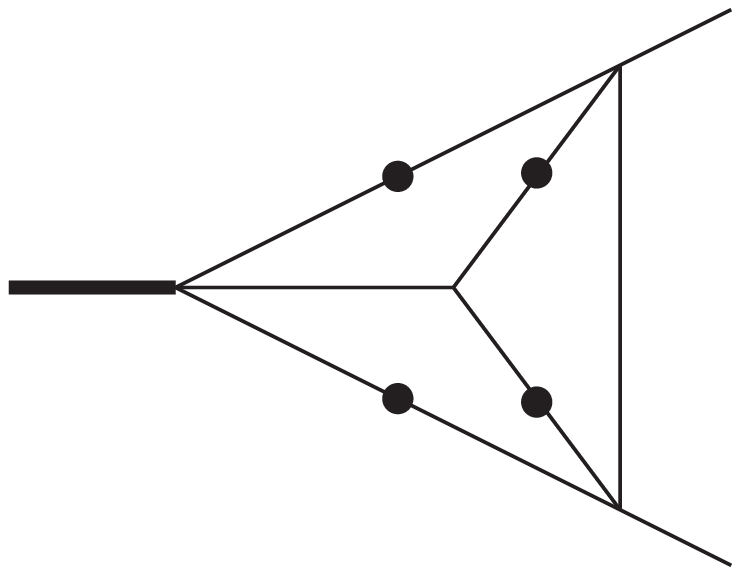}{6} \quad \figgraph{.35}{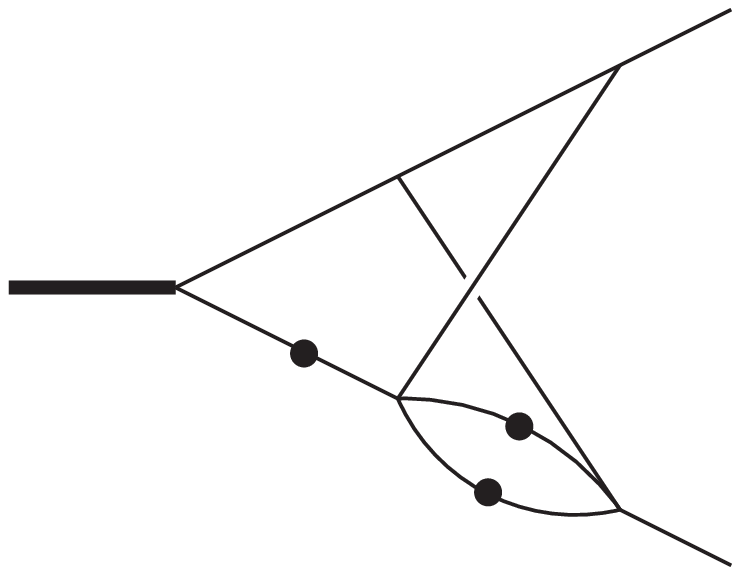}{6}
\\
&\figgraph{.35}{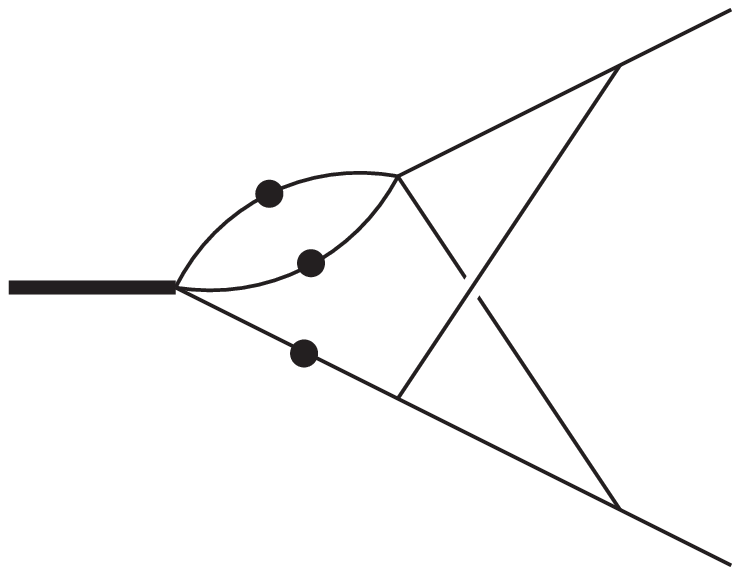}{6} \quad \figgraph{.35}{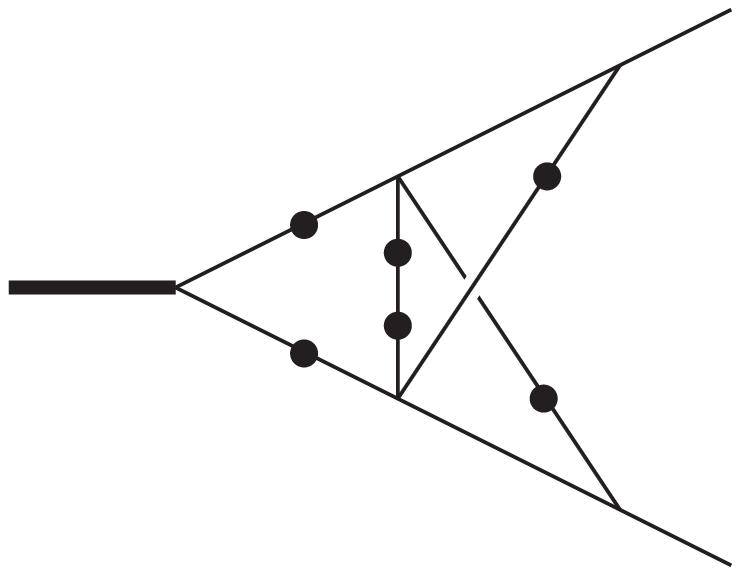}{8} \quad \figgraph{.35}{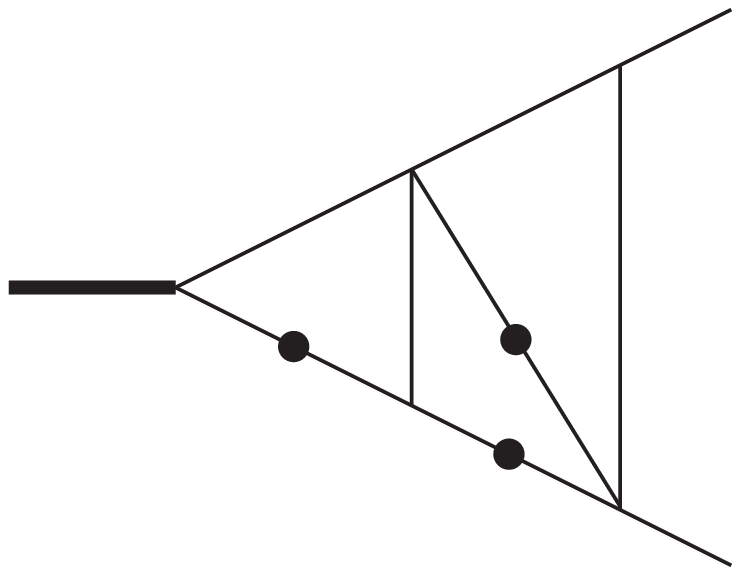}{6} \quad \figgraph{.35}{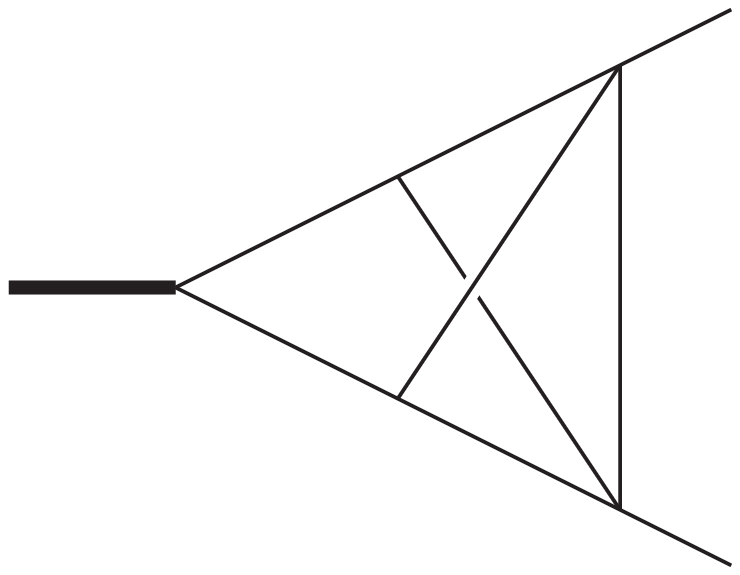}{4}
\\
&\figgraph{.35}{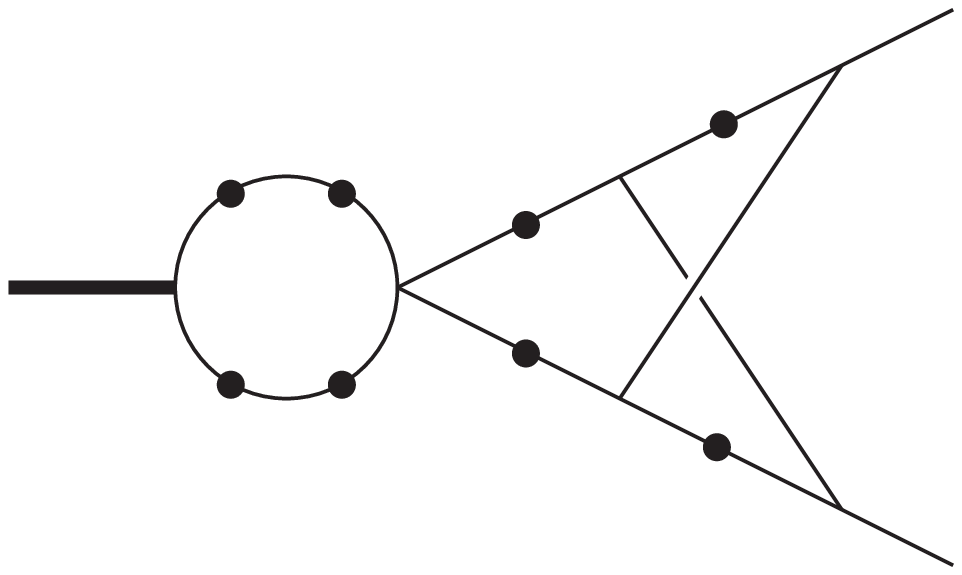}{8} \quad \figgraph{.35}{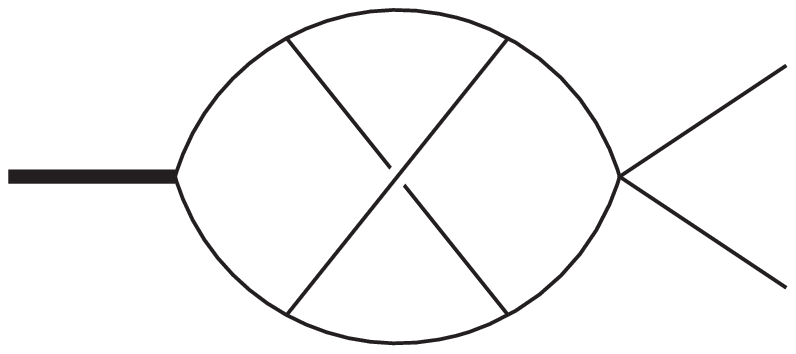}{4} \quad \figgraph{.35}{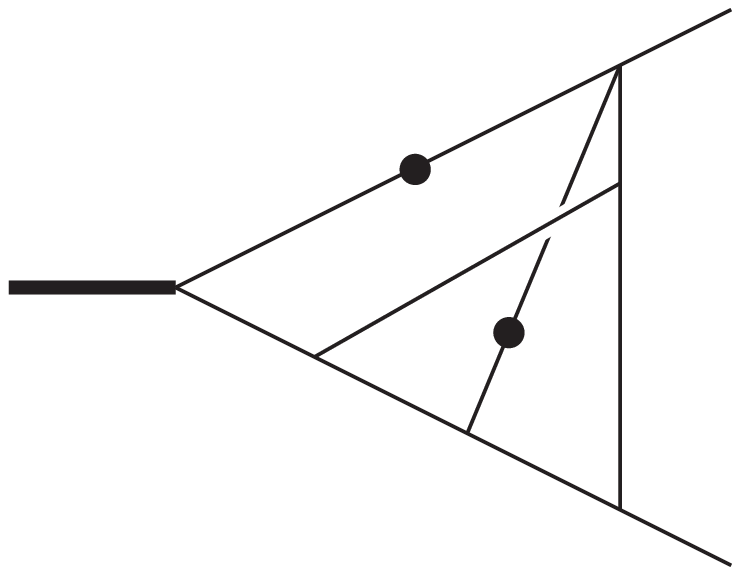}{6} \quad \figgraph{.35}{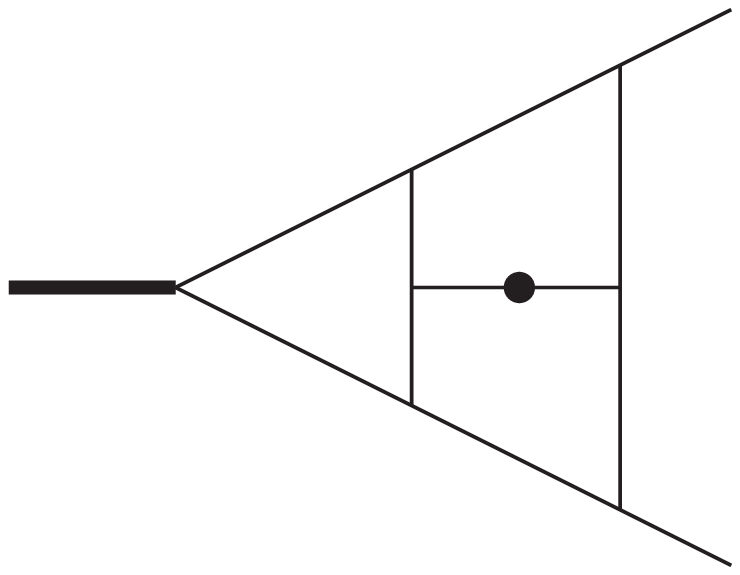}{6}
\\
&\figgraph{.35}{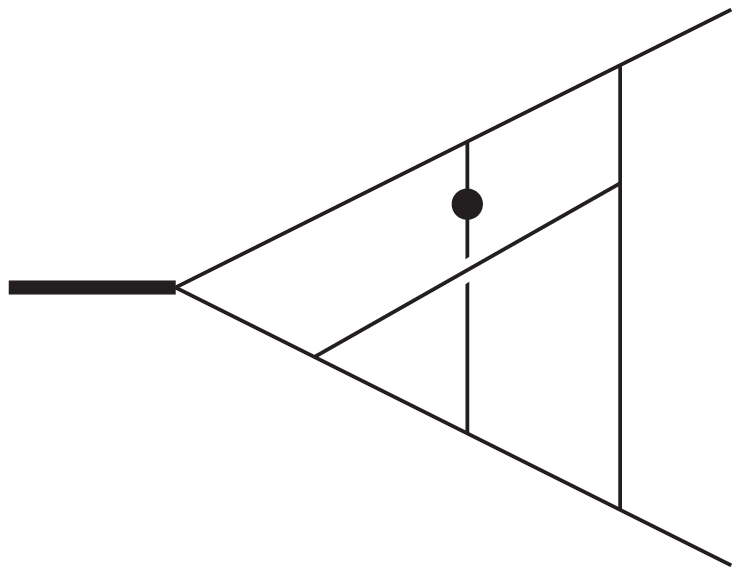}{6} \quad \figgraph{.35}{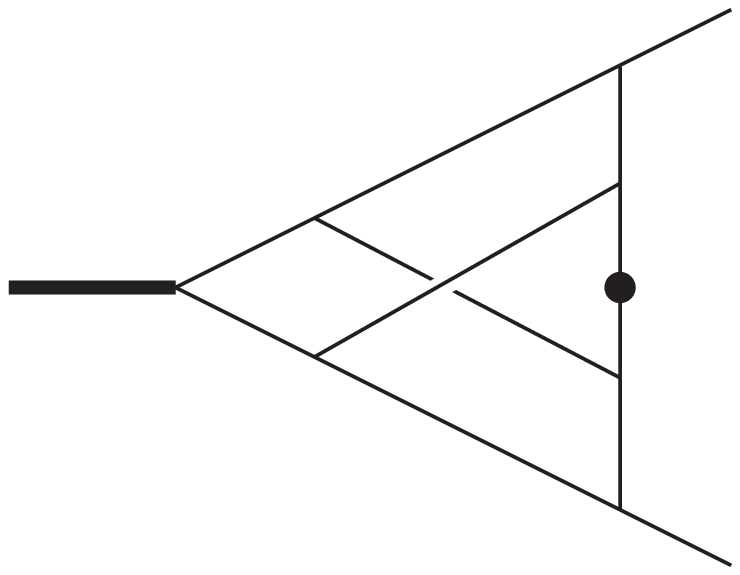}{6}
\end{align*}
\caption{The one-, two-, and three-loop finite form factor master integrals used in Sections \ref{sec:1Lffs}, \ref{sec:2Lffs}, and \ref{sec:3Lffs}
for $\mathcal{F}_1^q(\epsilon)$, $\mathcal{F}_1^g(\epsilon)$, $\mathcal{F}_2^q(\epsilon)$, $\mathcal{F}_2^g(\epsilon)$, $\mathcal{F}_3^q(\epsilon)$, and $\mathcal{F}_3^g(\epsilon)$.}
\label{fig:basis}
\end{figure}
Our calculation of the massless form factors is based on a variant of the method of dimension-shifts and dots recently introduced by us
in reference~\cite{vonManteuffel:2014qoa}. Its salient features can be summarized as follows.
Using Feynman diagrams, we express all relevant loop amplitudes in terms of scalar Feynman integrals which are then reduced to a set of basis integrals using integration-by-parts reductions.
For our basis integrals, we select Feynman integrals which are \emph{finite} in the $\epsilon \to 0$ limit.
Starting from the Feynman parameter representations of these integrals, we Taylor expand the integrands about $\epsilon = 0$. Finally, we use
modern analytical integration techniques to integrate all expansion coefficients in terms of zeta and multiple zeta values. 
We now proceed to describe how our calculations were carried out at a more technical level of detail.

As a first step, we generate all of the Feynman diagrams using {\tt Qgraf} \cite{Nogueira:1991ex} and compute the interferences required to extract the one-, two-, and three-loop form factors.
For this task, we use the QCD interference calculator built into the {\tt Reduze 2} program \cite{vonManteuffel:2012np,Studerus:2009ye,Bauer:2000cp,fermat}. 
The result of this computation is a linear combination of scalar loop integrals whose integrands have numerator insertions up to a certain maximal rank,
which are then mapped onto inverse propagators. We find that, in 't Hooft-Feynman gauge, there are up to two inverse propagators at one loop, 
up to four inverse propagators at two loops, and up to five inverse propagators at three loops.
Next, the loop integrals are systematically reduced to a minimal set of basis integrals, the so-called master integrals.
These reductions are obtained from integration by parts identities in $d$ spacetime
dimensions \cite{Tkachov:1981wb,Chetyrkin:1981qh} with a variant of Laporta's algorithm \cite{Laporta:2001dd}, 
as implemented in the {\tt Reduze 2} program.
At this point, we employ the well-known standard integral basis of corner
integrals in $4-2\epsilon$ dimensions:
one one-loop integral, four two-loop integrals, and twenty-two three-loop integrals.

In order to compute the master integrals using their Feynman parameter representations, we switch to a basis of \emph{finite master integrals}.\footnote{In this work, we chose to make use of scalar Feynman integrals without numerator insertions. 
However, it might be beneficial to take integrals with irreducible numerators into account, as this provides more finite integrals in a given number of dimensions.}
In previous work \cite{vonManteuffel:2014qoa}, we showed that one can make use of so-called quasi-finite master integrals, master integrals which have convergent Feynman parameter integrations but, 
potentially, a trivial overall $1/\epsilon$ divergence.
Here, we find it more convenient to work with truly finite integrals in order to make it manifestly visible which master integrals contribute to which $\epsilon$ poles of the bare form factors.
As explained in reference \cite{vonManteuffel:2014qoa} and Appendix~\ref{sec:finite}, such a finite basis can always be constructed in dimensional regularization for any multi-leg, 
multi-loop process, provided that there exists a Euclidean region which respects all kinematical constraints.

For each of the irreducible topologies, we enumerate finite integrals as described in Appendix~\ref{sec:finite}, using an automated job in the development version of {\tt Reduze 2}.
Out of many possible choices, we find it convenient
to keep the dotted graphical representations of our master integrals as symmetric as possible while aiming for masters possessing high maximal weights at leading order in $\epsilon$.
We discard sets of master integrals for which the reduced interferences contain spurious poles worse than $\epsilon^{-2 L}$ at $L$-loop order.
It is essential to avoid such spurious poles if one wishes to be able to read off by eye which master integrals contribute to which $\epsilon$ poles.
All other things being equal, we also find it natural to pick candidates which inhabit smaller spacetime dimensions and have fewer dots.
A systematic analysis of the first 100 finite integrals produced by our algorithm shows
that, for each sector, it is sufficient to use the lowest possible spacetime dimension ({\it i.e.} the spacetime dimension at which finite integrals first appear) to determine the highest possible maximal weight at leading order in $\epsilon$
consistent with our spurious pole veto.\footnote{Although we have no proof that one cannot do better, we were unable to find a candidate to cover 
the unique non-factorizable, three-point, eight-line integral topology which both had the highest maximal weight possible at leading order in $\epsilon$ (five) and resulted in reduced three-loop integrands without $\epsilon^{-7}$ poles.} 
In a few cases we picked non-minimal spacetime dimensions in order to allow for symmetric dottings.
The results of our analysis are depicted in Figure \ref{fig:basis}.

The next step is to relate the traditional integral basis to the basis of finite integrals using dimension shifts
and integration by parts relations, see Section 2.3 of reference~\cite{vonManteuffel:2014qoa} for details.
For this purpose, we must solve reduction identities for integrals besides the ones needed for the interferences.
Despite the fact that some of our finite integrals carry a rather large number
of dots, these reductions are moderate in computational complexity when compared
to the reductions required for the form factor Feynman diagrams. For the convenience of the reader, our arXiv submission includes the relevant lists of replacement rules required to pass from the standard Reduze 2
bases of corner integrals to the finite integral bases employed in this work (see \file{BasisToFinite.m}).

At this stage, we have reexpressed the one-, two-, and three-loop form factors as linear combinations of master integrals whose Feynman parameter integral representations converge in the $\epsilon \to 0$ limit.
If our basis integrals satisfy certain technical requirements, we can apply modern, highly-automated, analytical integration algorithms to compute the Taylor expansion about $\epsilon = 0$ for our master integrals.
The {\HyperInt} package \cite{Panzer:2014caa}, written in {\Maple}, provides powerful routines for the evaluation of Euclidean linearly reducible Feynman integrals\footnote{Roughly speaking, a Feynman integral 
is linearly reducible if its Feynman parametrization can be integrated out parameter-by-parameter in terms of multiple polylogarithms with rational arguments.} which possess convergent Feynman parameter representations.
The package implements the algorithms presented in \cite{Brown:2008um,Brown:2009ta} and all underlying principles are discussed in \cite{Panzer:2015ida}. In \cite{Panzer:2014gra}, 
it was pointed out that all three-loop form factor integrals happen to be linearly reducible. This non-obvious fact implies that one can use the {\HyperInt} approach to compute the $\epsilon$-expansion of the three-loop form factors. 
Even though the lower-loop results have been known to all orders in $\epsilon$ for some time, we use {\HyperInt} to evaluate the $\epsilon$-expansion of the form factors to weight eight in all cases. 
At one loop, this requires an $\epsilon$ expansion of the form factors up to $\mathcal{O}\left(\epsilon^6\right)$,
 at two loops, this requires an $\epsilon$ expansion of the form factors up to $\mathcal{O}\left(\epsilon^4\right)$, 
 and, at three loops, this requires an $\epsilon$ expansion of the form factors up to $\mathcal{O}\left(\epsilon^2\right)$. In the following three sections, we summarize our results.

\section{One-Loop Form Factors}
\label{sec:1Lffs}
For the bare one-loop form factors we find
\begin{align}
\label{eq:1Lffq}
\mathcal{F}_1^q(\epsilon) &=\casimir{C_F} \left\{\pole{\frac{1}{\epsilon^2}}\left[a_1 \dimgraph{ff1a_2_3}{6}\right]\right\}
\end{align}
\begin{align}
\label{eq:1Lffg}
\mathcal{F}_1^g(\epsilon) &=\casimir{C_A} \left\{\pole{\frac{1}{\epsilon^2}}\left[b_1 \dimgraph{ff1a_2_3}{6}\right]\right\},
\end{align}
where we have abbreviated the rational functions of $\epsilon$ as
\begin{align}
\label{eq:1Lffqcoeff}
a_1 &= \frac{-2 + \epsilon-2 \epsilon^2}{1-\epsilon}\\
\label{eq:1Lffgcoeff}
b_1 &= \frac{-2 (1-3 \epsilon+2 \epsilon^2+\epsilon^3)}{(1-\epsilon)^2}.
\end{align}
Here and in the following sections, the notation is such that both the graphically represented master integrals and the abbreviated rational
coefficient functions are finite at $\epsilon = 0$ and possess a Taylor series expansion starting at $\mathcal{O}\left(\epsilon^0\right)$, recall the form of \ Eq.\ \eqref{eq:miff1a3}.
This notation makes all divergences completely explicit and shows which integral topologies contribute to specific $\epsilon$ poles of the form factors.

At the one-loop level, the Casimir scaling property is reflected in the fact that 
$a_1$ is equal to $b_1$ in the $\epsilon \to 0$ limit.
Expanding Eqs.\ \eqref{eq:1Lffq} and \eqref{eq:1Lffg} to $\mathcal{O}\left(\epsilon^6\right)$, 
we find complete agreement with the results of references \cite{Gehrmann:2010ue} and \cite{Gehrmann:2010tu}. Further details of our analysis of the one-loop quark and gluon form factors are available online at arXiv.org 
in the ancillary files \file{Fq.m} and \file{Fg.m}.

\section{Two-Loop Form Factors}
\label{sec:2Lffs}
For the bare two-loop form factors we find
\begin{align}
\label{eq:2Lffq}
\mathcal{F}_2^q(\epsilon) &=\casimir{C_F^2} \left\{\pole{\frac{1}{\epsilon^4}} \left[c_1 \dimgraph{ff2a_4_15}{6} + c_2 \dimgraph{ff2a_3_22}{8}\right]
+\pole{\frac{1}{\epsilon^3}} \left[c_3 \dimgraph{ff2a_4_58}{10}\right]+\pole{\frac{1}{\epsilon}} \left[c_4 \dimgraph{ff2b_6_63}{8}\right]\right\}
\nonumber\\[-1pt] &
+ \casimir{C_F C_A} \left\{\pole{\frac{1}{\epsilon^4}} \left[c_5 \dimgraph{ff2a_3_22}{8}+c_6 \dimgraph{ff2a_4_58}{10}\right]+\pole{\frac{1}{\epsilon}} \left[c_7 \dimgraph{ff2b_6_63}{8}\right]\right\}
\nonumber\\[-1pt] &
+ \casimir{C_F N_f} \left\{\pole{\frac{1}{\epsilon^3}} \left[c_8 \dimgraph{ff2a_4_58}{10}\right]\right\}
\end{align}
\begin{align}
\label{eq:2Lffg}
\mathcal{F}_2^g(\epsilon) &=\casimir{C_A^2} \left\{\pole{\frac{1}{\epsilon^4}} \left[d_1 \dimgraph{ff2a_4_15}{6}+d_2 \dimgraph{ff2a_3_22}{8}+d_3 \dimgraph{ff2a_4_58}{10}\right]
+\pole{\frac{1}{\epsilon}} \left[d_4 \dimgraph{ff2b_6_63}{8}\right]\right\}
\nonumber\\[-1pt] &
+ \casimir{C_A N_f} \left\{\pole{\frac{1}{\epsilon^3}} \left[d_5 \dimgraph{ff2a_3_22}{8}+d_6 \dimgraph{ff2a_4_58}{10}\right]+d_7 \dimgraph{ff2b_6_63}{8}\right\}
\nonumber\\[-1pt] &
+ \casimir{C_F N_f} \left\{\pole{\frac{1}{\epsilon^2}} \left[d_8 \dimgraph{ff2a_3_22}{8}+d_9 \dimgraph{ff2a_4_58}{10}\right]+d_{10} \dimgraph{ff2b_6_63}{8}\right\}.
\end{align}
At this order in QCD perturbation theory, the Casimir scaling property can no longer be seen by eye; for example, the leading infrared divergences in the quark form factor exponentiate \cite{Frenkel:1976bj}
and the $C_F^2$ color structure can therefore not contribute to the two-loop quark anomalous dimension at all. What can be deduced from Eqs.\ \eqref{eq:2Lffq} and \eqref{eq:2Lffg}, 
however, is that the finite two-loop non-planar form factor integral in the above
cannot contribute to the two-loop quark and gluon cusp anomalous dimensions; it cannot contribute to the $\epsilon^{-2}$ pole terms of either form factor because
it is finite and has prefactors which, at worst, diverge like $\epsilon^{-1}$ in the $\epsilon \to 0$ limit. Expanding Eqs.\ \eqref{eq:2Lffq} and \eqref{eq:2Lffg} to $\mathcal{O}\left(\epsilon^4\right)$, 
we find complete agreement with the results of references \cite{Gehrmann:2010ue} and \cite{Gehrmann:2010tu}. Further details of our analysis of the two-loop quark and gluon form factors are available online at arXiv.org 
in the ancillary files \file{Fq.m} and \file{Fg.m}.

\section{Three-Loop Form Factors}
\label{sec:3Lffs}
\begin{figure}
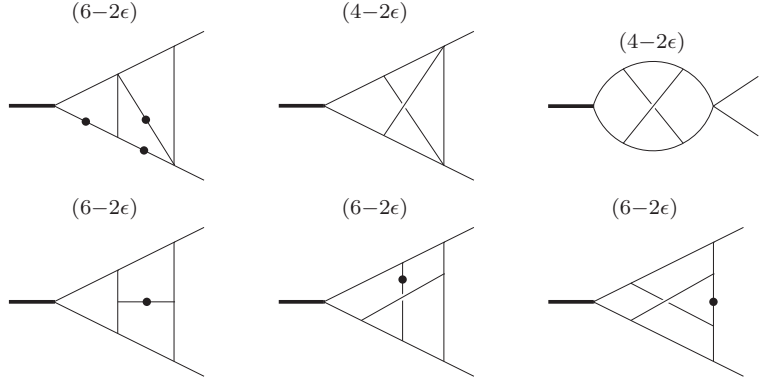

\centering
\begin{align*}
&\figgraph{.35}{ff3a_7_758}{6} \qquad \figgraph{.35}{ff3b_7_1722}{4} \qquad \figgraph{.35}{ff3b_8_2750}{4}
\\
&\figgraph{.35}{ff3a_9_1790}{6} \qquad \figgraph{.35}{ff3b_9_1790}{6} \qquad \figgraph{.35}{ff3c_9_1015}{6}
\end{align*}
\caption{The finite three-loop form factor master integrals in $\mathcal{F}_3^q\left(\epsilon\right)$ and $\mathcal{F}_3^g\left(\epsilon\right)$ (Eqs.\ \eqref{eq:3Lffq} and \eqref{eq:3Lffg})
which do not contribute to the three-loop cusp anomalous dimensions.}
\label{fig:noncuspintegrals}
\end{figure}

For the bare three-loop form factors we find
\begin{align}
\label{eq:3Lffq}
\mathcal{F}_3^q(\epsilon) &=
\casimir{C_F^3} \left\{\pole{\frac{1}{\epsilon^6}} \left[e_1 \dimgraph{ff3a_4_172}{10}+e_2 \dimgraph{ff3a_5_662}{8}+e_3 \dimgraph{ff3a_5_158}{6}+e_4 \dimgraph{ff3a_6_2695}{6}
\right.
\right.
\nonumber\\[-1pt] &
\left.
\left.
+e_5 \dimgraph{ff3a_6_1433}{10}\right]+\pole{\frac{1}{\epsilon^5}} \left[e_6 \dimgraph{ff3a_5_412}{10}+e_7 \dimgraph{ff3a_5_433}{8}+e_8 \dimgraph{ff3a_6_1683}{10}\right]
\right.
\nonumber\\[-1pt] &
\left.
+\pole{\frac{1}{\epsilon^4}} \left[e_9 \dimgraph{ff3a_6_429}{6}\right]+\pole{\frac{1}{\epsilon^3}} \left[e_{10} \dimgraph{ff3a_6_691}{8}+e_{11} \dimgraph{ff3a_6_444}{6}+e_{12} \dimgraph{ff3b_7_1770}{6}
\right.
\right.
\nonumber\\[-1pt] &
\left.
\left.
+e_{13} \dimgraph{ff3b_7_1780}{6}+e_{14} \dimgraph{ff3b_7_1766}{8}+e_{15} \dimgraph{ff3c_8_2959}{8}\right]+\pole{\frac{1}{\epsilon^2}} \left[e_{16} \dimgraph{ff3b_8_1662}{6}\right]
\right.
\nonumber\\[-1pt] &
\left.
+\pole{\frac{1}{\epsilon}} \left[e_{17} \dimgraph{ff3a_7_758}{6}+e_{18} \dimgraph{ff3b_7_1722}{4}+e_{19} \dimgraph{ff3b_8_2750}{4}+e_{20} \dimgraph{ff3a_9_1790}{6}+e_{21} \dimgraph{ff3b_9_1790}{6}
\right.
\right.
\nonumber\\[-1pt] &
\left.
\left.
+e_{22} \dimgraph{ff3c_9_1015}{6}\right]\right\}
\nonumber\\[5pt] & \hspace{-6.2ex}
+ \casimir{C_F^2 C_A} \left\{\pole{\frac{1}{\epsilon^6}} \left[e_{23} \dimgraph{ff3a_4_172}{10}+e_{24} \dimgraph{ff3a_5_662}{8}+e_{25} \dimgraph{ff3a_5_158}{6}+e_{26} \dimgraph{ff3a_5_412}{10}
\right.
\right.
\nonumber\\[-1pt] &
\left.
\left.
+e_{27} \dimgraph{ff3a_6_1683}{10}+e_{28} \dimgraph{ff3a_6_1433}{10}\right]+\pole{\frac{1}{\epsilon^5}} \left[e_{29} \dimgraph{ff3a_5_433}{8}\right]+\pole{\frac{1}{\epsilon^4}} \left[e_{30} \dimgraph{ff3a_6_429}{6}\right]
\right.
\nonumber\\[-1pt] &
\left.
+\pole{\frac{1}{\epsilon^3}} \left[e_{31} \dimgraph{ff3a_6_691}{8}+e_{32} \dimgraph{ff3a_6_444}{6}+e_{33} \dimgraph{ff3b_7_1770}{6}+e_{34} \dimgraph{ff3b_7_1780}{6}+e_{35} \dimgraph{ff3b_7_1766}{8}
\right.
\right.
\nonumber\\[-1pt] &
\left.
\left.
+e_{36} \dimgraph{ff3c_8_2959}{8}\right]+\pole{\frac{1}{\epsilon^2}} \left[e_{37} \dimgraph{ff3b_8_1662}{6}\right]+\pole{\frac{1}{\epsilon}} \left[e_{38} \dimgraph{ff3a_7_758}{6}+e_{39} \dimgraph{ff3b_7_1722}{4}
\right.
\right.
\nonumber\\[-1pt] &
\left.
\left.
+e_{40} \dimgraph{ff3b_8_2750}{4}+e_{41} \dimgraph{ff3a_9_1790}{6}+e_{42} \dimgraph{ff3b_9_1790}{6}+e_{43} \dimgraph{ff3c_9_1015}{6}\right]\right\}
\nonumber\\[5pt] & \hspace{-6.2ex}
+ \casimir{C_F C_A^2} \left\{\pole{\frac{1}{\epsilon^6}} \left[e_{44} \dimgraph{ff3a_4_172}{10}+e_{45} \dimgraph{ff3a_5_662}{8}+e_{46} \dimgraph{ff3a_5_158}{6}+e_{47} \dimgraph{ff3a_5_412}{10}
\right.
\right.
\nonumber\\[-1pt] &
\left.
\left.
+e_{48} \dimgraph{ff3a_5_433}{8}+e_{49} \dimgraph{ff3a_6_1433}{10}\right]+\pole{\frac{1}{\epsilon^4}} \left[e_{50} \dimgraph{ff3a_6_429}{6}\right]+\pole{\frac{1}{\epsilon^3}} \left[e_{51} \dimgraph{ff3a_6_444}{6}
\right.
\right.
\nonumber\\[-1pt] &
\left.
\left.
+e_{52} \dimgraph{ff3b_7_1770}{6}+e_{53} \dimgraph{ff3b_7_1780}{6}+e_{54} \dimgraph{ff3b_7_1766}{8}\right]+\pole{\frac{1}{\epsilon^2}} \left[e_{55} \dimgraph{ff3b_8_1662}{6}\right]
\right.
\nonumber\\[-1pt] &
\left.
+\pole{\frac{1}{\epsilon}} \left[e_{56} \dimgraph{ff3a_7_758}{6}+e_{57} \dimgraph{ff3b_7_1722}{4}+e_{58} \dimgraph{ff3a_9_1790}{6}+e_{59} \dimgraph{ff3b_9_1790}{6}+e_{60} \dimgraph{ff3c_9_1015}{6}\right]\right\}
\nonumber\\[5pt] & \hspace{-6.2ex}
+ \casimir{C_F^2 N_f} \left\{\pole{\frac{1}{\epsilon^5}} \left[e_{61} \dimgraph{ff3a_5_412}{10}+e_{62} \dimgraph{ff3a_5_433}{8}+e_{63} \dimgraph{ff3a_6_1683}{10}+e_{64} \dimgraph{ff3a_6_1433}{10}\right]
\right.
\nonumber\\[-1pt] &
\left.
+\pole{\frac{1}{\epsilon^4}} \left[e_{65} \dimgraph{ff3a_4_172}{10}\right]+\pole{\frac{1}{\epsilon^3}} \left[e_{66} \dimgraph{ff3a_6_429}{6}\right]+\pole{\frac{1}{\epsilon^2}} \left[e_{67} \dimgraph{ff3a_6_691}{8}
+e_{68} \dimgraph{ff3b_7_1770}{6}\right]\right\}
\nonumber\\[5pt] & \hspace{-6.2ex}
+ \casimir{C_F C_A N_f} \left\{\pole{\frac{1}{\epsilon^5}} \left[e_{69} \dimgraph{ff3a_4_172}{10}+e_{70} \dimgraph{ff3a_5_412}{10}+e_{71} \dimgraph{ff3a_5_433}{8}+e_{72} \dimgraph{ff3a_6_1433}{10}\right]
\right.
\nonumber\\[-1pt] &
\left.
+\pole{\frac{1}{\epsilon^3}} \left[e_{73} \dimgraph{ff3a_6_429}{6}\right]+\pole{\frac{1}{\epsilon^2}} \left[e_{74} \dimgraph{ff3a_6_444}{6}+e_{75} \dimgraph{ff3b_7_1770}{6}\right]\right\}
\nonumber\\[5pt] & \hspace{-6.2ex}
+ \casimir{C_F N_f^2} \left\{\pole{\frac{1}{\epsilon^4}}\left[e_{76} \dimgraph{ff3a_6_1433}{10}\right]\right\}
\nonumber\\[5pt] & \hspace{-6.2ex}
+ \casimir{\frac{d_{abc}d_{abc}}{N_c} N_{q\gamma}} \left\{\pole{\frac{1}{\epsilon^5}} \left[e_{77} \dimgraph{ff3a_4_172}{10}+e_{78} \dimgraph{ff3a_6_1433}{10}\right]+\pole{\frac{1}{\epsilon^4}}  \left[e_{79} \dimgraph{ff3a_5_433}{8}\right]
\right.
\nonumber\\[-1pt] &
\left.
+\pole{\frac{1}{\epsilon^3}} \left[e_{80} \dimgraph{ff3a_5_662}{8}+e_{81} \dimgraph{ff3a_5_158}{6}+e_{82} \dimgraph{ff3a_5_412}{10}+e_{83} \dimgraph{ff3a_6_429}{6}\right]
\right.
\nonumber\\[-1pt] &
\left.
+\pole{\frac{1}{\epsilon^2}} \left[e_{84} \dimgraph{ff3a_6_444}{6}+e_{85} \dimgraph{ff3b_7_1770}{6}+e_{86} \dimgraph{ff3b_7_1780}{6}+e_{87} \dimgraph{ff3b_7_1766}{8}\right]
\right.
\nonumber\\[-1pt] &
\left.
+\pole{\frac{1}{\epsilon}} \left[e_{88} \dimgraph{ff3b_8_1662}{6}\right]+\left[e_{89} \dimgraph{ff3b_7_1722}{4}+e_{90} \dimgraph{ff3a_9_1790}{6}+e_{91} \dimgraph{ff3b_9_1790}{6}\right]
\right.
\nonumber\\[-1pt] &
\left.
+\pole{\epsilon} \left[e_{92} \dimgraph{ff3a_7_758}{6}\right]\right\}
\end{align}
\begin{align}
\label{eq:3Lffg}
\mathcal{F}_3^g(\epsilon) &=
\casimir{C_A^3} \left\{\pole{\frac{1}{\epsilon^6}} \left[f_1 \dimgraph{ff3a_4_172}{10}+f_2 \dimgraph{ff3a_5_662}{8}+f_3 \dimgraph{ff3a_5_158}{6}+f_4 \dimgraph{ff3a_5_412}{10}+f_5 \dimgraph{ff3a_5_433}{8}
\right.
\right.
\nonumber\\[-1pt] &
\left.
\left.
+f_6 \dimgraph{ff3a_6_2695}{6}+f_7 \dimgraph{ff3a_6_1683}{10}+f_8 \dimgraph{ff3a_6_1433}{10}\right]+\pole{\frac{1}{\epsilon^4}} \left[f_9 \dimgraph{ff3a_6_429}{6}\right]
\right.
\nonumber\\[-1pt] &
\left.
+\pole{\frac{1}{\epsilon^3}} \left[f_{10} \dimgraph{ff3a_6_691}{8}+f_{11} \dimgraph{ff3a_6_444}{6}+f_{12} \dimgraph{ff3b_7_1770}{6}+f_{13} \dimgraph{ff3b_7_1780}{6}+f_{14} \dimgraph{ff3b_7_1766}{8}
\right.
\right.
\nonumber\\[-1pt] &
\left.
\left.
+f_{15} \dimgraph{ff3c_8_2959}{8}\right]+\pole{\frac{1}{\epsilon^2}} \left[f_{16} \dimgraph{ff3b_8_1662}{6}\right]+\pole{\frac{1}{\epsilon}} \left[f_{17} \dimgraph{ff3a_7_758}{6}+f_{18} \dimgraph{ff3b_7_1722}{4}
\right.
\right.
\nonumber\\[-1pt] &
\left.
\left.
+f_{19} \dimgraph{ff3b_8_2750}{4}+f_{20}\dimgraph{ff3a_9_1790}{6}+f_{21} \dimgraph{ff3b_9_1790}{6}\right]\right\}
\nonumber\\[5pt] & \hspace{-6.2ex}
+ \casimir{C_A^2 N_f} \left\{\pole{\frac{1}{\epsilon^5}} \left[f_{22} \dimgraph{ff3a_4_172}{10}+f_{23} \dimgraph{ff3a_5_662}{8}+f_{24} \dimgraph{ff3a_5_158}{6}+f_{25} \dimgraph{ff3a_5_412}{10}
\right.
\right.
\nonumber\\[-1pt] &
\left.
\left.
+f_{26} \dimgraph{ff3a_5_433}{8}+f_{27} \dimgraph{ff3a_6_1683}{10}+f_{28} \dimgraph{ff3a_6_1433}{10}\right]+\pole{\frac{1}{\epsilon^3}} \left[f_{29} \dimgraph{ff3a_6_429}{6}\right]
\right.
\nonumber\\[-1pt] &
\left.
+\pole{\frac{1}{\epsilon^2}} \left[f_{30} \dimgraph{ff3a_6_691}{8}+f_{31} \dimgraph{ff3a_6_444}{6}+f_{32} \dimgraph{ff3b_7_1770}{6}+f_{33} \dimgraph{ff3b_7_1780}{6}+f_{34} \dimgraph{ff3b_7_1766}{8}
\right.
\right.
\nonumber\\[-1pt] &
\left.
\left.
+f_{35} \dimgraph{ff3c_8_2959}{8}\right]+\pole{\frac{1}{\epsilon}} \left[f_{36} \dimgraph{ff3b_8_1662}{6}\right]+\left[f_{37} \dimgraph{ff3a_7_758}{6}+f_{38} \dimgraph{ff3b_7_1722}{4}\right.
\right.
\nonumber\\[-1pt] &
\left.
\left.+f_{39} \dimgraph{ff3b_8_2750}{4}+f_{40} \dimgraph{ff3a_9_1790}{6}+f_{41} \dimgraph{ff3b_9_1790}{6}+f_{42} \dimgraph{ff3c_9_1015}{6}\right]\right\}
\nonumber\\[5pt] & \hspace{-6.2ex}
+ \casimir{C_A C_F N_f} \left\{\pole{\frac{1}{\epsilon^5}} \left[f_{43} \dimgraph{ff3a_4_172}{10}+f_{44} \dimgraph{ff3a_5_662}{8}+f_{45} \dimgraph{ff3a_5_158}{6}+f_{46} \dimgraph{ff3a_5_433}{8}
\right.
\right.
\nonumber\\[-1pt] &
\left.
\left.
+f_{47} \dimgraph{ff3a_6_1433}{10}\right]+\pole{\frac{1}{\epsilon^4}} \left[f_{48} \dimgraph{ff3a_5_412}{10}+f_{49} \dimgraph{ff3a_6_1683}{10}\right]+\pole{\frac{1}{\epsilon^2}} \left[f_{50} \dimgraph{ff3a_6_691}{8}
\right.
\right.
\nonumber\\[-1pt] &
\left.
\left.
+f_{51} \dimgraph{ff3a_6_444}{6}+f_{52} \dimgraph{ff3b_7_1770}{6}+f_{53} \dimgraph{ff3b_7_1780}{6}+f_{54} \dimgraph{ff3b_7_1766}{8}+f_{55} \dimgraph{ff3c_8_2959}{8}
\right]
\right.
\nonumber\\[-1pt] &
\left.
+\pole{\frac{1}{\epsilon}} \left[f_{56} \dimgraph{ff3a_6_429}{6}+f_{57} \dimgraph{ff3b_8_1662}{6}\right]+\left[f_{58} \dimgraph{ff3a_7_758}{6}+f_{59} \dimgraph{ff3b_7_1722}{4}+f_{60} \dimgraph{ff3b_8_2750}{4}\right.
\right.
\nonumber\\[-1pt] &
\left.
\left. +f_{61} \dimgraph{ff3a_9_1790}{6}+f_{62} \dimgraph{ff3b_9_1790}{6}+f_{63} \dimgraph{ff3c_9_1015}{6}\right]\right\}
\nonumber\\[5pt] & \hspace{-6.2ex}
+\casimir{C_F^2 N_f} \left\{\pole{\frac{1}{\epsilon^4}} \left[f_{64} \dimgraph{ff3a_5_433}{8}+f_{65} \dimgraph{ff3a_6_1433}{10}\right]+\pole{\frac{1}{\epsilon^3}} \left[f_{66} \dimgraph{ff3a_4_172}{10}+f_{67} \dimgraph{ff3a_5_158}{6}
\right.
\right.
\nonumber\\[-1pt] &
\left.
\left.
+f_{68} \dimgraph{ff3a_5_412}{10}\right]+\pole{\frac{1}{\epsilon^2}} \left[f_{69} \dimgraph{ff3a_5_662}{8}+f_{70} \dimgraph{ff3a_6_444}{6}+f_{71} \dimgraph{ff3b_7_1780}{6}\right]
\right.
\nonumber\\[-1pt] &
\left.
+\pole{\frac{1}{\epsilon}} \left[f_{72} \dimgraph{ff3b_7_1770}{6}+f_{73} \dimgraph{ff3b_7_1766}{8}\right] + \left[f_{74} \dimgraph{ff3a_6_429}{6}+f_{75} \dimgraph{ff3a_9_1790}{6}+f_{76} \dimgraph{ff3c_9_1015}{6}\right]
\right.
\nonumber\\[-1pt] &
\left.
+\epsilon \left[f_{77} \dimgraph{ff3a_7_758}{6}+f_{78} \dimgraph{ff3b_7_1722}{4}\right]\right\}
\nonumber\\[5pt] & \hspace{-6.2ex}
+ \casimir{C_A N_f^2} \left\{\pole{\frac{1}{\epsilon^4}}\left[f_{79} \dimgraph{ff3a_4_172}{10}+f_{80} \dimgraph{ff3a_5_158}{6}+f_{81} \dimgraph{ff3a_5_412}{10}\right]+\pole{\frac{1}{\epsilon^2}}\left[f_{82} \dimgraph{ff3a_6_1433}{10}\right]
\right.
\nonumber\\[-1pt] &
\left.
+\pole{\frac{1}{\epsilon}}\left[f_{83} \dimgraph{ff3b_7_1780}{6}\right]\right\}
\nonumber\\[5pt] & \hspace{-6.2ex}
+ \casimir{C_F N_f^2} \left\{\pole{\frac{1}{\epsilon^3}}\left[f_{84} \dimgraph{ff3a_4_172}{10}+f_{85} \dimgraph{ff3a_5_412}{10}\right]+\pole{\frac{1}{\epsilon^2}}\left[f_{86} \dimgraph{ff3a_5_158}{6}\right]
\right.
\nonumber\\[-1pt] &
\left.
+\pole{\frac{1}{\epsilon}}\left[f_{87} \dimgraph{ff3b_7_1780}{6}\right]
\right\}.
\end{align}
In total, six of the twenty-two finite three-loop form factor master integrals in Eqs.\ \eqref{eq:3Lffq} and \eqref{eq:3Lffg} above turn out not to contribute to the three-loop cusp anomalous dimensions (see Figure \ref{fig:noncuspintegrals}). 
These include, in particular, all of the most complicated, nine-line, finite three-loop masters. Expanding Eqs.\ \eqref{eq:3Lffq} and \eqref{eq:3Lffg} to $\mathcal{O}\left(\epsilon^2\right)$, 
we find complete agreement with the results of references \cite{Gehrmann:2010ue} and \cite{Gehrmann:2010tu}. Further details of our analysis of the three-loop quark and gluon form factors are available online at arXiv.org 
in the ancillary files \file{Fq.m} and \file{Fg.m}.

\section{Towards the Four-Loop Cusp Anomalous Dimensions}
\label{sec:cusps}
As has long been known, a convenient way to calculate the cusp anomalous dimensions to four loops is to perform the four-loop quark and gluon form factor calculations and then extract the four-loop cusp anomalous dimensions
from the first non-trivial poles in the parameter of dimensional regularization. 
The $\epsilon^{-8} - \epsilon^{-3}$ poles at four loops are predicted by the renormalization group equations satisfied by the form factors, in terms of known lower-loop results,
and the $\epsilon^{-2}$ poles may be used to extract the four-loop cusp anomalous dimensions (see {\it e.g.} reference~\cite{Moch:2005tm} for details).\footnote{In reference \cite{Gehrmann:2010ue},
the one-, two-, and three-loop form factors are given to sufficiently high orders in the $\epsilon$ expansion for the purposes of this analysis.}

Normally, for dimensionally-regulated, $L$-loop bare amplitudes in quantum field theory, which
can be expressed in terms of multiple zeta values or, more generally, multiple polylogarithms, one expects
contributions of, at most, weight $2 L + n$ in the Laurent expansion coefficient of order $n$.\footnote{We are not aware of any proof that this will always be the case. 
However, to the best of our knowledge, this weight bound turns out to hold in every explicit higher-order calculation performed to date.}
In fact, given our experience at lower loop orders, it is natural to hope
that the four-loop cusp anomalous dimensions are rational linear combinations of the numbers $1$, $\zeta_2$, $\zeta_3$, $\zeta_2^2$, $\zeta_2 \zeta_3$, $\zeta_5$, $\zeta_2^3$, and $\zeta_3^2$.
If we further assume that there are no cancellations of spurious constants between different master integrals, we
conclude that integrals which contain other transcendental constants at leading order in $\epsilon$ should not contribute to the four-loop cusp anomalous dimensions.
For illustration, we have chosen the finite, non-planar twelve-line integral\footnote{Here we use the convention $\displaystyle \zeta_{5,3}=\sum_{0<n<m} n^{-3} m^{-5}$.}
\begin{align}\label{eq:4L}
	\dimgraph{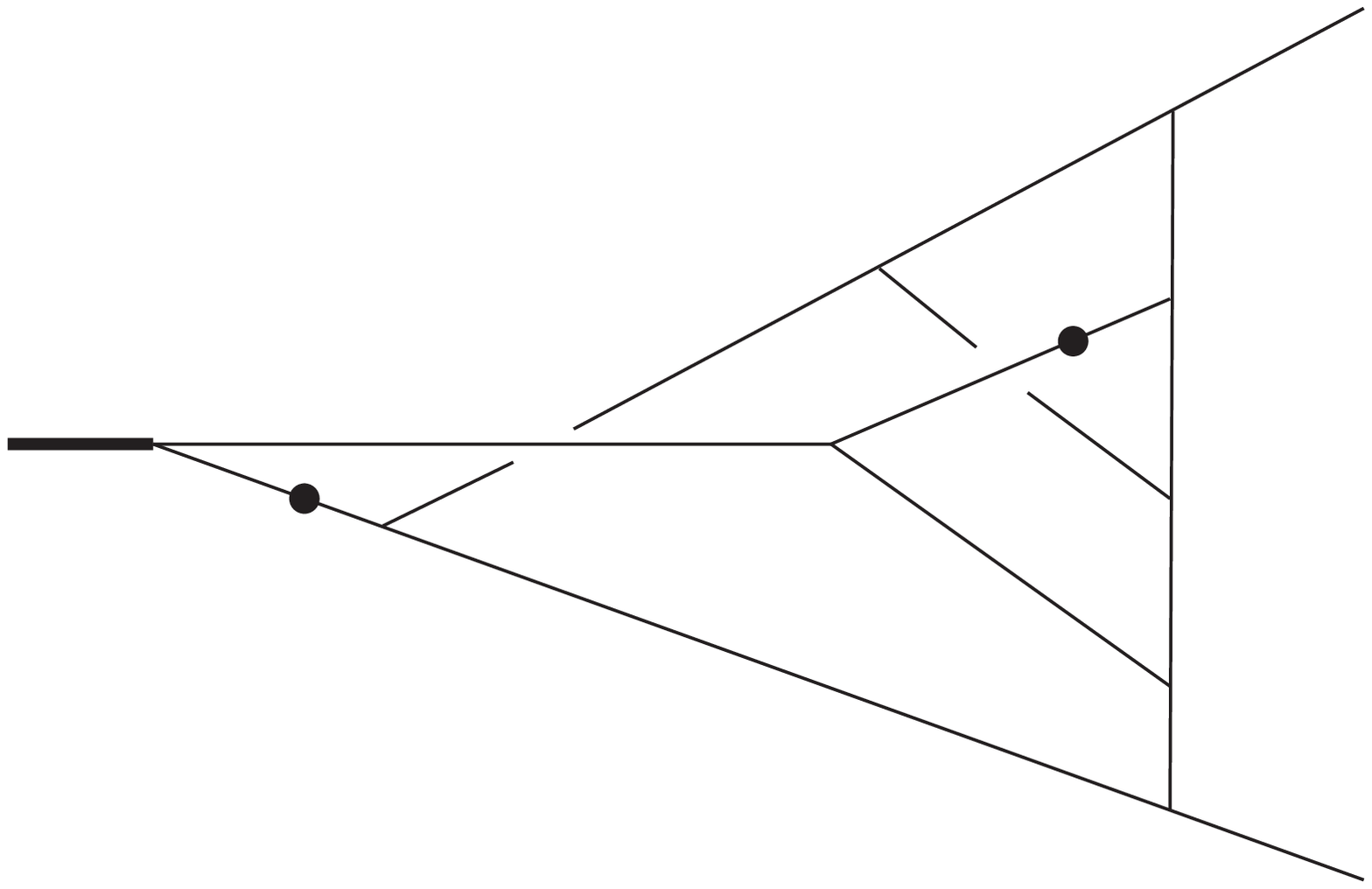}{6} 
	& =
	\tfrac{18}{5} \zeta^2_2 \zeta_3 - 5 \zeta_2\zeta_5
	+\Big(
		24 \zeta_2 \zeta_3
		+20 \zeta_5
		-\tfrac{188}{105} \zeta_2^3
		-17 \zeta_3^2
		+9 \zeta_2^2 \zeta_3 
	\nonumber\\&
		-47 \zeta_2 \zeta_5 
		-21 \zeta_7
		+\tfrac{6883}{2100} \zeta_2^4
		+\tfrac{49}{2} \zeta_2 \zeta_3^2
		+\tfrac{1}{2} \zeta_3 \zeta_5
		-9 \zeta_{5,3}
	\Big)\epsilon
	+\mathcal{O}\left(\epsilon^2\right),
\end{align}
which is of pure weight seven at leading order in $\epsilon$ and should thus not contribute to the four-loop cusp anomalous dimensions in our approach.
We computed \eqref{eq:4L} analytically with the programs in the ancillary files as described in appendix~\ref{sec:hyperint} and checked our result to four significant digits numerically using {\tt FIESTA 3} \cite{Smirnov:2013eza},
a program which provides a highly-automated framework for the numerical evaluation of Feynman integrals via sector decomposition \cite{Binoth:2000ps}.

In fact, we see at four loops indications of a pattern analogous to what was observed at two and three loops. The simplest integral topologies ({\it e.g.} those with many bubble insertions) give rise to finite Feynman integrals
which are likely to contribute to all or almost all poles in the form factors, due to the fact that they have relatively small maximal weights at leading order in the $\epsilon$ expansion. 
On the other hand, already at nine lines we found several examples of integral topologies which are not likely to be relevant to the calculation of the $\epsilon^{-2}$ poles in the form factors; in these sectors, we have identified
finite integrals which have maximal weights of seven or even eight at leading order in $\epsilon$.

Out of all four-loop form factor integral topologies, only fifteen fail {\HyperInt}'s test for linear reducibility out of the box and ten of these are top-level topologies. Altogether, this means that roughly 90\% of all relevant
Feynman integrals can be accessed using the methods implemented in {\HyperInt}. Given our experience so far, 
it seems possible that at least the set of finite Feynman integrals which contribute to the four-loop cusp anomalous dimensions could be computable analytically using the technology that we have developed.
Let us point out, however, that the integration by parts reductions required for the four-loop form factors in QCD are challenging to calculate.
In order to overcome the limitations of available programs, an implementation of the new
integration by parts reduction technique described in reference \cite{vonManteuffel:2014ixa} is under development.

\section{Conclusions}
\label{sec:conclusions}
In this paper, we studied the bare quark and gluon form factors in massless QCD using a recently-developed approach to the computation of virtual corrections in quantum field theory. Using integration by parts reduction,
we found compact expressions for the one-, two-, and three-loop form factors in terms of linearly reducible, finite master integrals. 
Linear reducibility and finiteness together guarantee that, at least in principle, the deterministic integration algorithms implemented in {\HyperInt}
can immediately be applied to analytically compute the $\epsilon$ expansion of these master integrals to the order required.
We evaluated all master integrals to sufficiently high order and then checked all of the literature on the subject.
At the three-loop level, our analysis constitutes the first complete independent analytical check on the
published higher-order results.

The extraction of the cusp anomalous dimensions at $L$ loop order requires knowledge of the $L$-loop form factors expanded through to $\mathcal{O}\left(\epsilon^{-2}\right)$.
With our choice of finite master integrals, only a subset of the full set of master integrals contributes to these poles in $\epsilon$.
At the three-loop level, we have shown explicitly that none of the most complicated (nine-line) form factor integrals contribute to the cusp anomalous dimensions.
Although our analysis of the four-loop form factor integrals is still ongoing,
we have found that the finite integrals in many sectors seem to follow an analogous pattern and are not expected to contribute to the four-loop cusp anomalous dimensions.
Once the necessary interferences and integration by parts reductions become available,
it will be possible to algebraically check all four-loop integral topologies to see whether or not they contribute to the $\epsilon^{-2}$ pole terms.

We encountered fifteen irreducible four-loop form factor topologies for which a straightforward test of linear reducibility failed.
These integrals are interesting in their own right, even if it turns out that they do not contribute to the four-loop cusp anomalous dimensions.
Investigating them is worthwhile because they will likely be relevant to the finite parts of the four-loop form factors 
and studying them could quite possibly lead to the development of even more powerful methods for the analytical evaluation of Feynman integrals.
For example, it might be that these integrals become linearly reducible once one makes a suitable change of variables. 
Non-trivial examples of this phenomenon were reported recently in \cite[section~4]{Panzer:2014gra} and \cite[section~2.1]{Panzer:2014fla},
but it is not yet completely clear how to determine whether or not a given topology which fails to be linearly reducible can be treated in this way.

It may be that, at least initially, some of the four-loop form factor integrals defeat all known analytical methods. 
Whether or not an exact determination of all basis integrals is possible to sufficiently high order in the $\epsilon$ expansion, our method of dimension shifts and dots allows one to make use of public software packages for sector decomposition
(principally {\tt sector\_decomposition} \cite{Bogner:2007cr}, {\tt FIESTA 3} \cite{Smirnov:2013eza}, and {\tt SecDec 3} \cite{Borowka:2015mxa}) 
in situations where it would almost certainly not be possible to proceed in the conventional approach utilized in {\it e.g.} reference \cite{Boels:2015yna}.
The analysis upon which this assertion is based will be discussed at length in a forthcoming publication on numerical applications of our method.

\section*{Acknowledgments}
We gratefully acknowledge Bas Tausk for bringing reference \cite{Tkachov:1996wh} and follow-up work to our attention and Cedric Studerus for discussions.
We are grateful to Johannes Schlenk for pointing out a misprint in the diagram
in \eqref{eq:4L} in the original version of this paper to us.
The work of RMS was supported by the European Research Council (ERC)
through the Advanced Grant EFT4LHC and Grant 647356 (CutLoops).
EP was supported by ERC Grant 257638 via the CNRS at the IHES and the Humboldt-Universit\"{a}t zu Berlin. Our figures were generated using {\tt Jaxodraw} \cite{Binosi:2003yf}, based on {\tt AxoDraw} \cite{Vermaseren:1994je}.

\appendix
\section{From Quasi-Finite Integrals to Finite Integrals}
\label{sec:finite}
In this appendix, we show that, in Euclidean kinematics, it is possible to pass from a basis of quasi-finite Feynman integrals to a basis of truly finite integrals, 
free of both infrared and ultraviolet divergences.\footnote{
Employing finite parametric integrals for Feynman integrals was proposed
in the Bernshtein-Tkachov approach \cite{Bernshtein:1972,Tkachov:1996wh}, long before the appearance of reference \cite{vonManteuffel:2014qoa}.
The method was subsequently worked out for the numerical evaluation of one-loop
and lower-point two-loop integrals, see \cite{Ferroglia:2002mz,Passarino:2006gv}.
However, it seems challenging to apply
to the general multi-loop case.}
Suppose we start with a basis of quasi-finite master integrals, $B$, where we allow for a shifted number of dimensions and higher powers of the given denominators (for existence and construction of this basis see \cite{vonManteuffel:2014qoa}).
Each of these integrals has convergent Feynman parameter integrations by definition but possibly an overall $1/\epsilon$ ultraviolet divergence.
Let us consider a quasi-finite integral $I\in B$ with such a divergence in the Gamma function prefactor of its Feynman parametrization.
In our normalization, this representation of our scalar, $L$-loop, $d$-dimensional, $N$-propagator Feynman integral $I$ is given by
\begin{align}
\label{eq:parametric-rep}
I &= \frac{\Gamma^L(d/2-1) \Gamma\big(\nu - L d/2\big) (-1)^{\nu}}{\prod_{k=1}^N \Gamma(\nu_k)} \times 
\nonumber \\
&\times\Bigg[ \prod_{j=1}^N \int_0^{\infty} {\rm d} x_j x_j^{\nu_j-1}\Bigg] \delta(1-x_N)\,\mathcal{U}^{\nu - (L+1)d/2} \mathcal{F}^{-\nu + L d/2},
\end{align}
where $\mathcal{U}$ and $\mathcal{F}$ are the Symanzik polynomials, $\nu_i$ denotes an integer propagator power and $\nu = \sum_{i=1}^N \nu_i$ is the sum of these multiplicities.
For $I$ to have an overall ultraviolet divergence, it must be the case that $\nu - L d/2$, the argument of the overall Gamma function prefactor in the above, is less than or equal to zero in the $\epsilon \to 0$ limit.

Since this is the case by assumption, we multiply the integrand of $I$ by $\mathcal{F}/\mathcal{F}$ to obtain
\begin{align}
\label{eq:I}
I =& \frac{\Gamma^L((d+2)/2-1) \Gamma\big(\nu+L+1 - L (d+2)/2\big) (-1)^{\nu}}{\big(d/2-1\big)^L \big(\nu - L d/2\big)\prod_{k=1}^N \Gamma(\nu_k)} \times
\nonumber\\
&\times \Bigg[ \prod_{j=1}^N \int_0^{\infty} {\rm d} x_j x_j^{\nu_j-1}\Bigg] \delta(1-x_N)\, \mathcal{U}^{\nu+L+1 - (L+1)(d+2)/2} \mathcal{F}^{-(\nu+L+1) + L (d+2)/2} \mathcal{F}.
\end{align}
Now, recall that $\mathcal{F}=\sum_m c_m \prod_{k=1}^N x^{m_k}$ is a homogeneous polynomial of degree $L + 1$ in the Feynman parameters with coefficients $c_m$ determined by the kinematic invariants.
Therefore,
\begin{equation}
	I
	=
	\sum_m
	I_m
	\frac{
		(-1)^{L+1} c_m
	}{
		(d/2-1)^L
		(\nu- L d/2 )
	}
	\prod_{k=1}^N \frac{\Gamma(\nu_k+m_{k})}{\Gamma(\nu_k)}
\end{equation}
is a linear combination of the Feynman integrals $I_m$ in $d+2$ dimensions with propagator powers $\nu_k+m_{k}$ ($L+1$ additional dots) and an improved superficial degree of divergence $\nu+L+1 - L (d+2)/2 = (\nu - L d/2) + 1$.
Note that the convergence of the Feynman parameter integrations at $\epsilon=0$, which we imposed on $I$, 
continues to hold for each of the $I_m$ because, in an appropriately chosen Euclidean region where all coefficients $c_m$ are non-negative, $0\leq c_m \prod_k x_k^{m_{k}}\leq\mathcal{F}$ everywhere in the domain of integration.

Clearly, the steps outlined above can be repeated, producing a truly finite Feynman integral after some number of iterations.
By virtue of the linear independence of the set $B$ it must be possible to pick a suitable term in $\mathcal{F}$ at every step in the above
procedure to ensure that the new integral is independent of $B \setminus \{I\}$.
We can discard $I$ from our basis and replace it with a new, genuinely finite Feynman integral.
If necessary, this process can be repeated until only finite integrals remain in our basis.

Note that, in practice, it is not necessary to actually go through the explicit steps described above. Instead, since they constructively prove that a basis of finite master integrals does exist, 
it suffices to simply enumerate a relatively large number of integrals according to their dottings and dimension shifts, check which ones are finite, and then choose a convenient finite basis from amongst the available candidates. 
This can be done with existing integration by parts programs as described in \cite[section~2.5]{vonManteuffel:2014qoa}.

\section{Automated Feynman Integral Evaluations with {\HyperInt}}
\label{sec:hyperint}
One motivation for our recomputation of the form factor integrals is to confirm the results published before with an independent method. 
In particular, some of the highest weight contributions have so far only been inferred from a numerical computation \cite{Lee:2010ik} using the PSLQ algorithm \cite{PSLQ}.
We think that it is very important to enable other researchers to straightforwardly reproduce such results for themselves.
To this end, a key component of our arXiv submission is our computational framework for the evaluation of the form factor master integrals.
In the folder \file{MIintegration} of the ancillary files, we provide our complete setup for the expansion of the master integrals in a form which allows everyone with access to {\Maple} and appropriate computational resources to recompute them.
With a modest amount of additional effort, we expect that the computer codes we present may be modified for the automated computation of other kinds of linearly reducible Feynman integrals.

\subsection{Linear Reducibility}
To compute the master integrals, we expand the integrand of \eqref{eq:parametric-rep} in $\epsilon$ and use the algorithm described in reference \cite{Brown:2008um} to successively integrate out the Feynman parameters one-by-one. 
The prerequisite to apply this method is called \emph{linear reducibility} \cite{Brown:2009ta} and it puts constraints on the complexity of the Symanzik polynomials.
It was observed in \cite[Theorem~3.4]{Panzer:2014gra} that this criterion is fulfilled for all massless three-point three-loop integrals even in the case of arbitrary external momenta, 
hence it applies in particular to our special case where $p_1^2=p_2^2=0$.

\subsection{Practical Remarks}
We used {\HyperInt} \cite{Panzer:2014caa} and prepared several scripts to automate the integration of the master integrals. The whole computation proceeds as follows:
\begin{enumerate}
	\item The master integrals, specified in \texttt{Reduze} format, are parsed into {\Maple} format and bubbles are integrated out in terms of Gamma functions.\footnote{This simplification could in fact be skipped, 
	but it might become performance-critical at four loops.}

	\item For each remaining parametric integral \eqref{eq:parametric-rep}, the Symanzik polynomials $\mathcal{U}$ and $\mathcal{F}$ are analyzed through a polynomial reduction \cite[appendix~A]{Panzer:2014gra}. 
	From this, a linearly reducible integration order is constructed automatically.\footnote{This functionality is part of the 
	dev branch of \href{https://bitbucket.org/PanzerErik/hyperint/}{{\HyperInt} on Bitbucket} (\texttt{suggestIntegrationOrder}).}

	\item Individual terms of the $\epsilon$ expansion of the parametric integrals can now be computed separately, {\it i.e.}\ in parallel, and we provide full support for parallelized runs.

	\item All results of such integrations are finally combined with the appropriate Gamma prefactors to construct the desired $\epsilon$ expansions.
\end{enumerate}
In order to decrease the running time of the computations, we applied two optimizations:
\begin{enumerate}
	\item As recommended in the {\HyperInt} manual, the polynomial reduction is used to discard in advance contributions which are known to cancel (these ``restricted regularizations'' were introduced already in \cite{Brown:2008um}). 
	This speeds up the integration itself.

	\item After each integration step, we express the integrand (which depends on the remaining Feynman parameters) in a ``product-of-hyperlogarithms'' basis (see \cite[appendix~A.1]{Panzer:2014gra} and \cite{Panzer:2015ida}) of 
	linearly independent functions \cite[command~\texttt{fibrationBasis}]{Panzer:2014caa}. This ensures that no relations persist among the polylogarithms which make up the integrand. 
	While this increases the number of terms present during the first steps of integration, it reduces the expression size considerably at later stages once several variables have been integrated out and functions of higher transcendental
	weight start to make an appearance.

		Let us also point out that {\HyperInt} provides a command, \texttt{fibreIntegration}, for this combination of integration with projection onto a basis (see \file{integrate.mpl} in the ancillary files for its usage).
\end{enumerate}
In the ancillary files, we provide all of our scripts in the \file{MIintegration} folder. The \file{README} file gives detailed, simple instructions concerning the use of these programs.

\bibliographystyle{JHEP}
\bibliography{ff3l}

\end{document}